\documentclass[reprint,superscriptaddress,aps,prb,twocolumn]{revtex4-2}
\usepackage{setspace}
\usepackage{amsmath}
\usepackage{graphicx}
\usepackage{subfig}
\usepackage{ragged2e}
\usepackage{amsfonts}
\usepackage{amssymb}
\usepackage{epstopdf} 
\usepackage{xcolor}
\usepackage{verbatim}
\usepackage{soul}
\usepackage{comment}
\usepackage{placeins}
\usepackage{textcomp}
\usepackage{bm}
\usepackage{hyperref}

\DeclareGraphicsExtensions{.pdf,.eps,.png,.jpg,.mps} 

\usepackage{float}
\renewcommand{\figurename}{\textbf{Figure}}

\begin{document}

\title{Soliton pulse pairs at multiple colors in normal dispersion microresonators}

\author{Zhiquan Yuan$^{1,*}$, Maodong Gao$^{1,*}$, Yan Yu$^{1,*}$, Heming Wang$^{1,2,*}$, Warren Jin$^{2,3,*}$, Qing-Xin Ji$^{1}$, Avi Feshali$^{3}$, Mario Paniccia$^{3}$, John Bowers$^{2,\dagger}$, and Kerry Vahala$^{1,\dagger}$\\
$^1$T. J. Watson Laboratory of Applied Physics, California Institute of Technology, Pasadena, CA 91125, USA.\\
$^2$ECE Department, University of California Santa Barbara, Santa Barbara, CA 93106, USA.\\
$^3$Anello Photonics, Santa Clara, CA 95054, USA.\\
$^*$These authors contributed equally to this work. \\ 
$^\dagger$Corresponding authors: jbowers@ucsb.edu, vahala@caltech.edu}

\begin{abstract}
Soliton microcombs \cite{kippenberg2018dissipative} are helping to advance the miniaturization of a range of comb systems \cite{diddams2020optical}. These combs mode lock through the formation of short temporal pulses in anomalous dispersion resonators. Here, a new microcomb is demonstrated that mode locks through the formation of pulse pairs in normal-dispersion coupled-ring resonators. Unlike conventional microcombs, pulses in this system cannot exist alone, and instead must phase lock in pairs to form a bright soliton comb. Also, the pulses can form at recurring spectral windows and the pulses in each pair feature different optical spectra. This pairwise mode-locking modality extends to higher dimensions and we demonstrate 3-ring systems in which 3 pulses mode lock through alternating pairwise pulse coupling. The results are demonstrated using the new CMOS-foundry platform that has not previously produced bright solitons on account of its inherent normal dispersion \cite{jin2021hertz}. The ability to generate multi-color pulse pairs over multiple rings is an important new feature for microcombs. It can extend the concept of all-optical soliton buffers and memories \cite{Wabnitz_OL1993,Leo2010} to multiple storage rings that multiplex pulses with respect to soliton color and that are spatially addressable. The results also suggest a new platform for the study of quantum combs \cite{reimer2016generation, kues2017chip, kues2019quantum} and topological photonics \cite{Topo2014,RevModPhys.91.015006,tikan2022protected}.
\end{abstract}

\maketitle

\section{Introduction}

\begin{figure*}[t!]
\begin{centering}
\includegraphics[width=\linewidth]{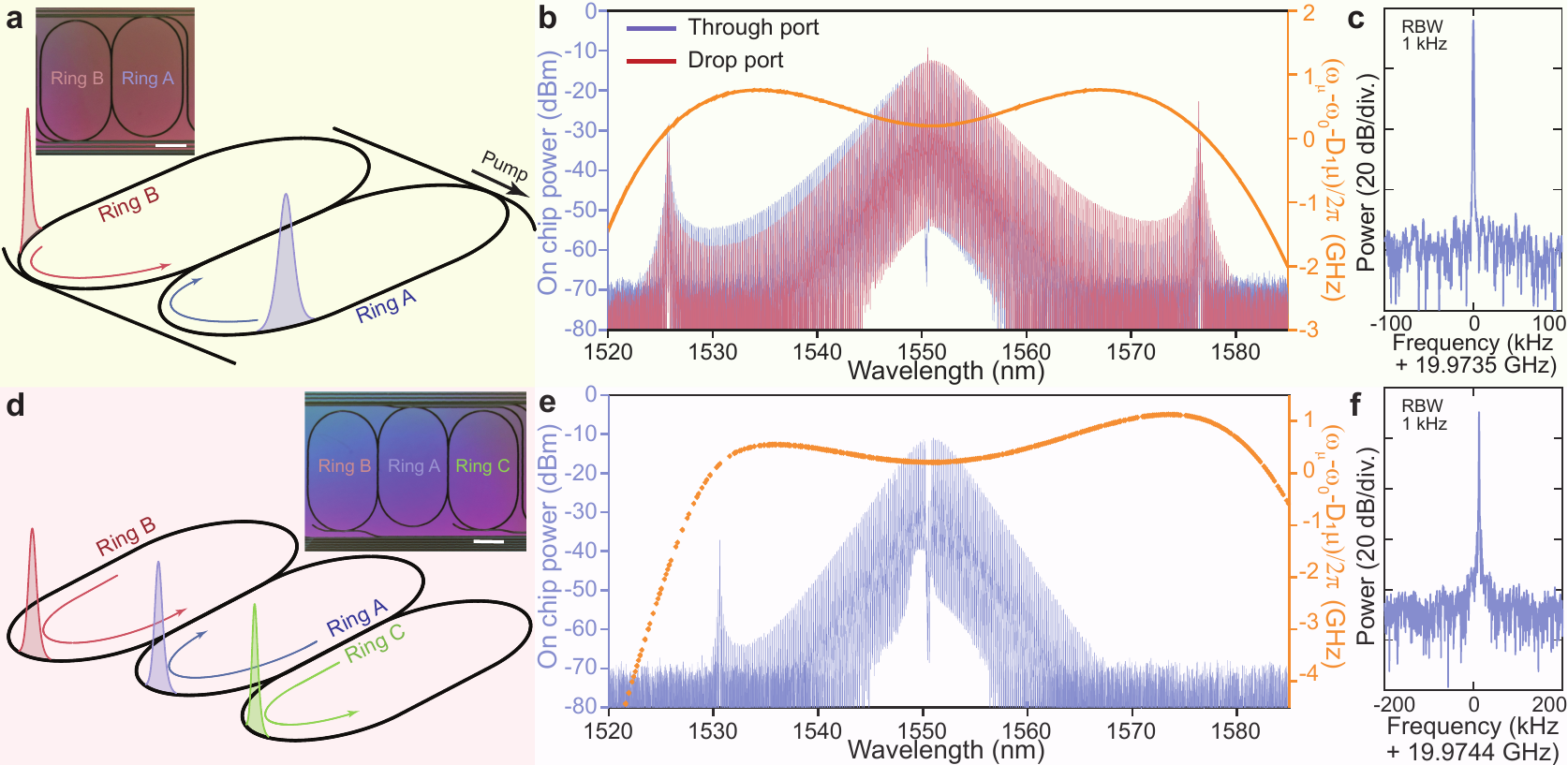}
\captionsetup{singlelinecheck=off, justification = RaggedRight}
\caption{{\bf Soliton pulse pair generation in two- and three-coupled-ring microresonators.}
\textbf{a} Schematic showing coherent pulse pairs that form a composite excitation.
Inset: Photomicrograph of the two-coupled-ring resonator used in the experiments. Rings A and B are indicated. The scale bar is 1 mm.
\textbf{b} Simultaneous measurement of optical spectra collected from the through port (pumping port) and drop port in the coupled-ring resonator of panel {\bf a}. The measured mode dispersion is also plotted (orange). Two dispersive waves are observed at spectral locations corresponding to the phase matching condition as indicated by the dispersion curve.
\textbf{c} Radio-frequency spectrum of microcomb beatnote (RBW: resolution bandwidth).
\textbf{d} Illustration of 3 pulse generation in a three-coupled-ring microresonator wherein pulses alternately pair.
Inset: Photomicrograph of the three-coupled-ring microresonator used in the experiments. The scale bar is 1 mm.
\textbf{e} Measurement of optical spectrum of the three pulse microcomb. The measured mode dispersion is also plotted (orange). 
\textbf{f} Radio frequency spectrum of the microcomb beatnote.
}
\label{Fig1}
\end{centering}
\end{figure*}

Microresonator solitons exist through a balance of optical nonlinearity and dispersion, which must be anomalous for bright soliton generation. Moreover, microresonators must feature high optical Q factors for low pump power operation of the resulting microcomb. While these challenges have been addressed at telecommunications wavelengths using a range of material systems \cite{kippenberg2018dissipative}, CMOS-foundry resonators do not yet support bright solitons as their waveguides feature normal dispersion \cite{jin2021hertz}. Furthermore, all resonators are dominated by normal dispersion at shorter wavelengths. For these reasons, there has been keen interest in developing methods to induce anomalous dispersion for bright soliton generation in systems that otherwise feature normal dispersion. Such methods have in common the engineering of dispersion through coupling of resonator mode families, including those associated with concentric resonator modes \cite{soltani2016enabling, kim2017dispersion}, polarization  \cite{lee2017towards} or transverse modes \cite{karpov2018}. 

Here, we engineer anomalous dispersion in CMOS-foundry resonators by partially-coupling resonators as illustrated in Fig. \ref{Fig1}a. This geometry introduces unusual new features to bright soliton generation. For example, spectra resembling single pulse microcombs form instead from pulse pairs as illustrated in Fig. \ref{Fig1}a. The pulse pairs circulate in a mirror-image fashion in the coupled rings to form coherent comb spectra (Fig. \ref{Fig1}b) with highly stable microwave beat notes (Fig. \ref{Fig1}c). The interaction of the pulses in the coupling section between the rings is shown to induce anomalous dispersion that compensates for the overall normal dispersion of each ring. This pairwise compensation spectrally recurs thereby opening multiple anomalous dispersion windows for the formation of multi-color soliton pairs. These windows can be engineered during resonator design. Furthermore, the spectral composition of each pulse in a pair is different. Fig. \ref{Fig1}b, for example, shows through-port and drop-port spectra that reflect the distinct spectral compositions of pulses in cavity A and cavity B of Fig. \ref{Fig1}a. This peculiar effect is also associated with Dirac solitons \cite{wang2020dirac} and it is shown that the 2-ring pulse pair represents a new embodiment of a Dirac soliton as the underlying dynamical equation (see Methods) resembles the nonlinear Dirac equation in $1+1$ dimensions. Pulse pairing is also extendable to higher-dimensional designs with additional normal dispersion rings. For example, in Fig. \ref{Fig1}d-f 3 pulses in 3 coupled rings alternately pair to compensate for the normal dispersion of each ring. 

\begin{figure*}[t!]
\begin{centering}
\includegraphics[width=\linewidth]{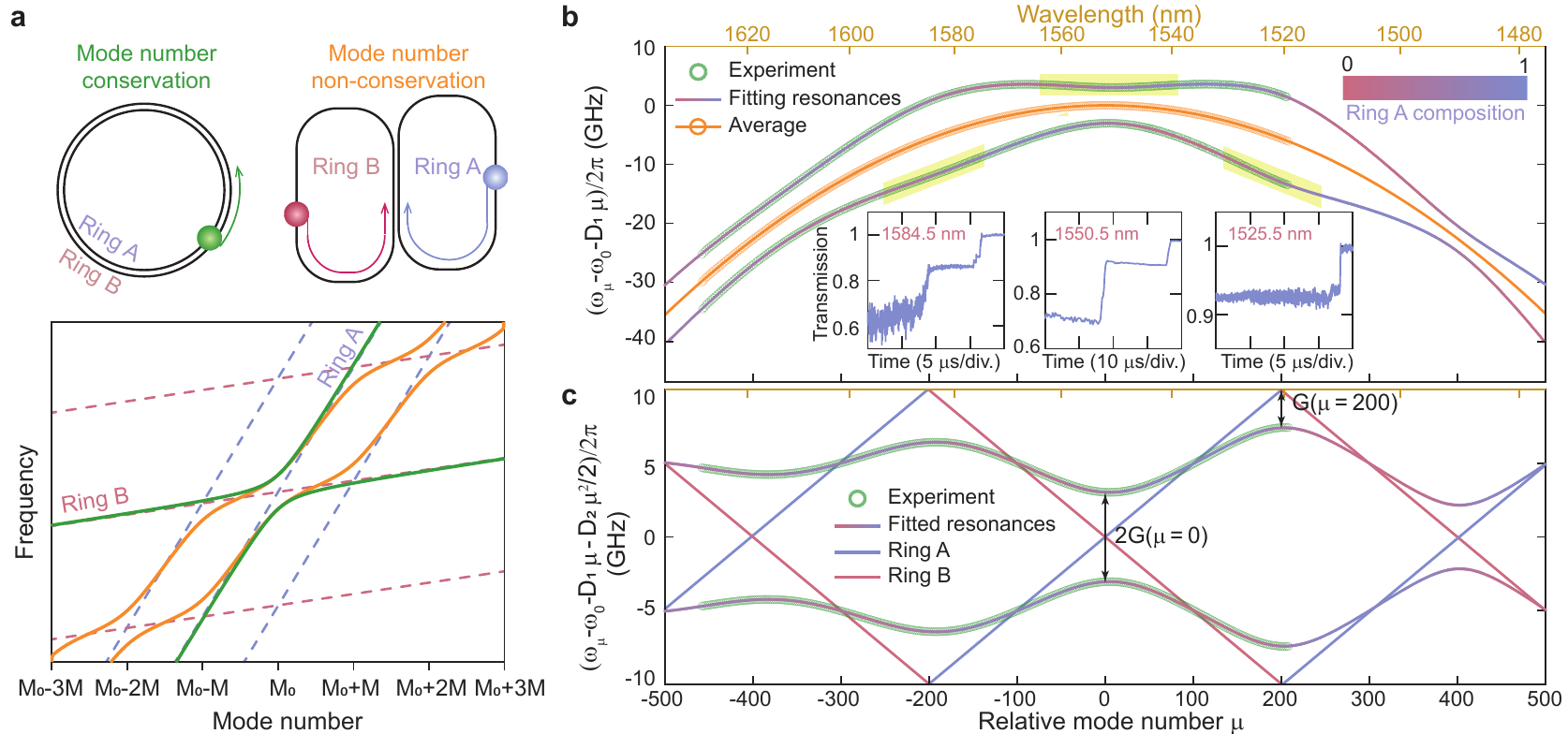}
\captionsetup{singlelinecheck=off, justification = RaggedRight}
\caption{{\bf Mode number non-conservation coupling and recurring bright soliton windows.}
\textbf{a} Illustration of inter-ring coupling with (without) mode number conservation. The top panel shows two different coupling schemes between two ring resonators with different FSRs. The left configuration possesses a continuous rotational symmetry that allows coupling only between modes with the same absolute mode (azimuthal) number (i.e., mode number is conserved). In this case, the coupling opens a gap at the mode crossing and creates two hybrid mode branches (green curves in the lower panel). Here the center blue and red dashed lines represent the resonance frequency of individual rings, and their intersection point corresponds to phase matching. 
In this work (top right panel), inter-ring mode coupling depends on the matching of resonance frequency instead of mode number (i.e., mode number is not conserved), so that the dispersion is strongly altered at all frequency degeneracies.
In the lower panel, frequency degeneracies are marked by crossings between the blue and red dashed lines, which still represent the dispersion of individual rings, but with the abscissa shifted by integer numbers as a result of spectral folding allowed by non-conservation of mode number. Compared to the first configuration, the dispersion curve of the coupled rings repeats itself every $2M$ modes, with $M=1/(2\epsilon)$ set by the length contrast of the rings.
\textbf{b} Measured frequency dispersion of the coupled resonator (green circles) versus relative mode number $\mu$. Here $D_1/(2\pi) = 19.9766$ GHz, and $\omega_0$ is chosen so that $\mu$ = 0 is at the crossing center (1552.3 nm).
Multiple anomalous dispersion windows appear around $\mu$ = $0$ and $400$ for the upper branch and $\mu$ = $-200$ and $200$ for the lower branch. The anomalous dispersion window near $\mu$ = $-200$, $0$ and $200$ have been highlighted. 
Solid curves are fittings and the color refers to the energy contribution from ring A (obtained from theoretical calculations).
The average of the upper and lower branch mode frequencies is plotted as orange circles and fitted by a second-order dispersion model (orange curve).
Inset: transmission observed when scanning a laser over resonances in the anomalous dispersion windows. Soliton steps are observed around $\mu$ = $-200$, $0$ and $200$.
\textbf{c} Measured relative frequency dispersion of the coupled resonator (green circles) versus relative mode number $\mu$. Here $D_2/(2\pi) = -283.0$ kHz, and other parameters are the same as panel \textbf{b}. Solid curves are the theoretical fittings described by Eq. (\ref{dispersion_gap}). 
Fitted mode frequency dispersion diagrams of the single rings without coupling are shown as red and blue lines.
}
\label{Fig2}
\end{centering}
\end{figure*}

In what follows, we first study the dispersion of this system and compare it to previous mode coupling methods. Experimental results including dispersion measurement and comb formation are then presented. Pairwise pulse formation is then studied in the time domain. Finally, because multi-pulse spectra in these systems resemble conventional single-pulse soliton spectra, it is convenient to resolve this ambiguity by denoting 2 and 3 ring systems as bipartite and tripartite soliton microcombs, respectively. The need for this nomenclature becomes clear by the demonstration of multiple pulse-pair states, including a 2 ring microcomb state containing 4 pulses that behaves as a 2-pulse soliton crystal, and a 3 ring state with 12 pulses that behaves as a 4-pulse soliton crystal.

\section{Recurring spectral windows}

Before addressing pulse pair propagation in the 2-ring and 3-ring systems, the conventional mode-family coupling approach is considered \cite{soltani2016enabling,lee2017towards,kim2017dispersion}. As a representative example, the case of a concentric resonator system is chosen as illustrated in the left panel of Fig. \ref{Fig2}a. The characteristics of this system are identical to other methods. First, a phase matching condition must be satisfied by a mode in each resonator such that the absolute mode number in each ring must be equal at the same optical frequency. This mode number determines the wavelength where soliton formation is possible. Second, the free-spectral-range values, FSR$_A$ and FSR$_B$, of the uncoupled mode families of ring A and ring B must be close in value compared to their average FSR = (FSR$_A$ + FSR$_B$)/2 so that phase matching occurs over a large number of modes. With these conditions satisfied, the resulting dispersion will be as illustrated schematically in the lower panel of Fig. \ref{Fig2}a (green curves). Comparison to the uncoupled dispersion curves (center dashed blue and red lines) shows that anomalous dispersion is possible for the upper mode family branch in the spectral vicinity of the phase matching mode number ${\rm M}_0$.

Next, consider the case where two rings are placed side-by-side and coupled together as illustrated in the right panel of Fig. \ref{Fig2}a. The two ring cavities differ only in length, with ring B slightly longer than ring A so that FSR$_A$$>$FSR$_B$. Considering the straight coupling section from a coupled-mode perspective, modes of the two rings will strongly couple if they have matching wavevectors (or equivalently, resonance frequencies), while there are no requirements on mode number matching of the rings (i.e., mode number is not conserved). In comparison to the concentric ring configuration, this dramatically modifies the dispersion relation as illustrated in Fig \ref{Fig2}a, where the orange curves give the resulting dispersion. Due to the loss of mode number conservation, inter-ring coupling pushes the resonance frequencies away from that of the individual rings (blue and red dashed lines) at all frequency degeneracies, so that recurring anomalous dispersion windows now appear in the spectrum. These result from spectral folding that occurs between the cavity resonances. As an aside, because mode number is not conserved, modeling of this dispersion proceeds differently relative to the standard coupled-mode family approach (see supplementary materials). 

\section{Dispersion measurements and soliton pulse pair generation}

\begin{figure*}[t!]
\begin{centering}
\includegraphics[width=\linewidth]{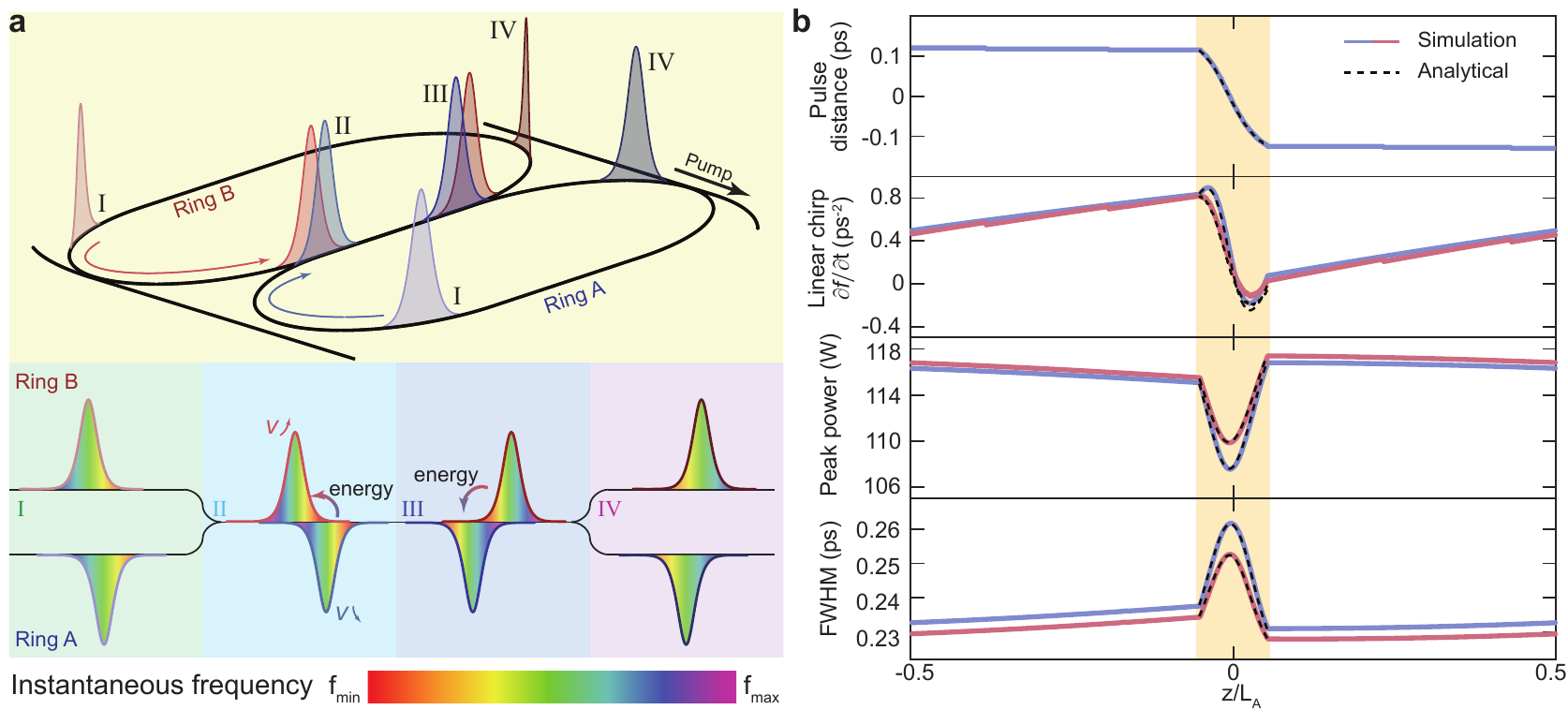}
\captionsetup{singlelinecheck=no, justification = RaggedRight}
\caption{{\bf Temporal evolution of the soliton pulse pair in the two-ring coupled resonator.}
\textbf{a} Upper panel: Illustration of the time evolution of the soliton pair inside the two rings during one round trip time. Lower panel: Snapshots of the pulses at different positions. In the non-coupled regions (I and IV), pulses accumulate positive chirp due to nonlinearity and normal dispersion of the waveguide. Pulse in ring A is leading in time at I due to shorter ring circumference. When the pulses enter the coupling region (II), the pulses exchange energy, which leads to relative position shifts as well as chirp compensation (III). The pulses exit the coupled region (IV) with position shifts and chirping compensated.
\textbf{b} Simulated pulse pair properties are plotted versus pulse position in each ring during one round trip.
The two rings are aligned at the coupling region center, and the surplus length in ring B is omitted in the figure. The yellow shaded area represents the coupling region.
The quantities are, from top to bottom: pulse timing difference (pulse center-to-center), linear chirp, peak power, and full width at half maximum. The blue (red) lines represent simulation results for the pulse in ring A (B). The dashed lines are analytical results from a linear coupling model (see Methods), and are consistent with simulation results.
}
\label{Fig3}
\end{centering}
\end{figure*}

\begin{figure*}[t!]
\begin{centering}
\includegraphics[width=\linewidth]{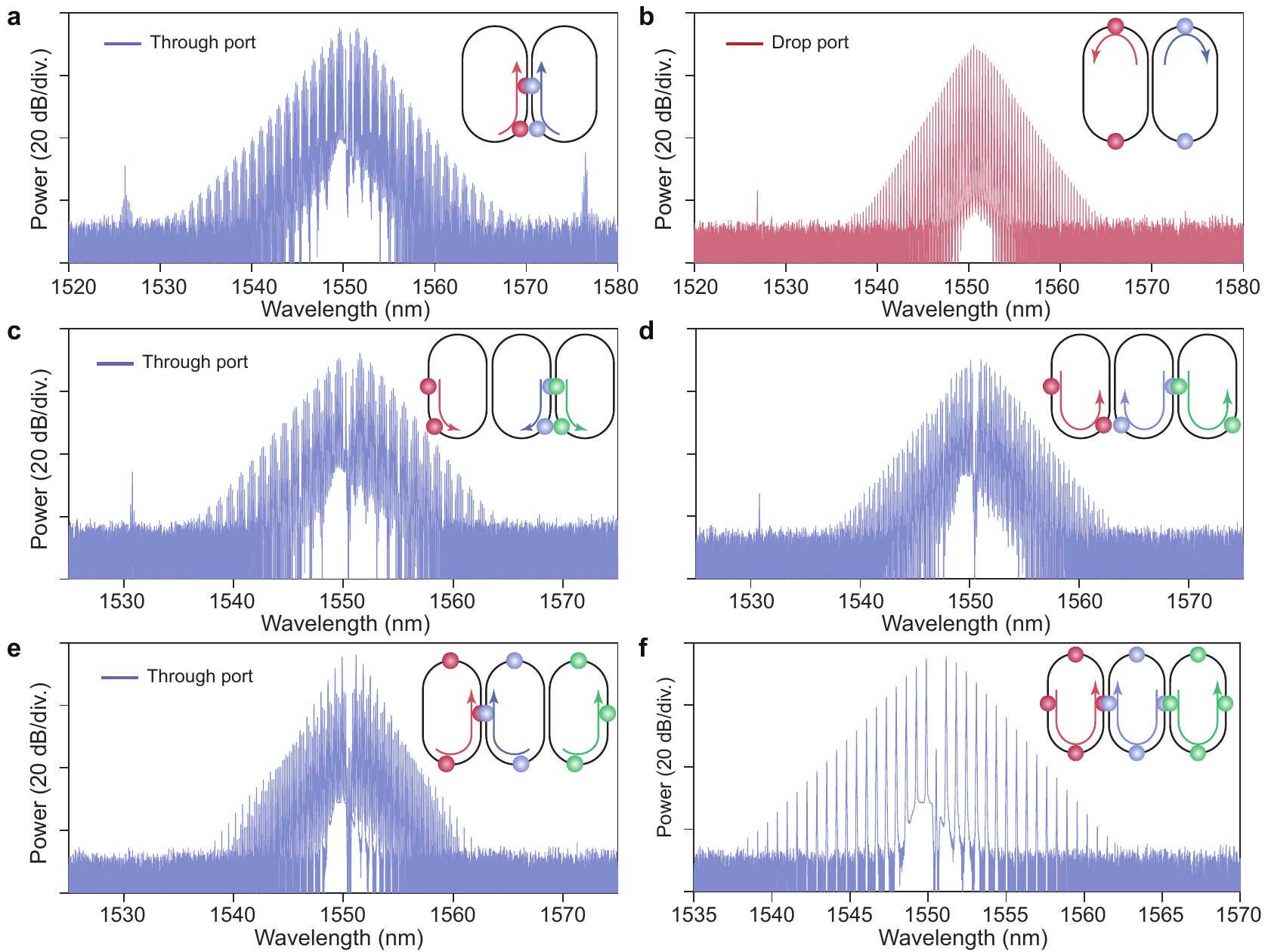}
\captionsetup{singlelinecheck=no, justification = RaggedRight}
\caption{{\bf Observation of bipartite and tripartite multi soliton states in two- and three-coupled-ring microresonators.}
\textbf{a, b} Optical spectra of bipartite two-soliton states with different relative soliton positions. The state in panel {\bf b} is a two-soliton crystal state.
Insets: relative position of the two solitons inside each microresonator.
\textbf{c, d} Through port optical spectra of tripartite two-soliton states with different relative positions. Inset: relative position of the two solitons inside each microresonator.
\textbf{e} Through port optical spectrum of a tripartite three-soliton state. Inset: relative position of the three solitons inside each microresonator.
\textbf{f} Through port optical spectrum of a tripartite four-soliton crystal. Inset: relative position of the four solitons inside each microresonator.}
\label{Fig4}
\end{centering}
\end{figure*}

The coupled resonators in both 2-ring and 3-ring geometries consist of thin Si$_3$N$_4$ single-mode racetrack waveguide resonators with the same cross sections (see Extended Data Fig. \ref{ExFig2}b). Bus waveguides are provided for external coupling. Optical images of 2-ring and 3-ring coupled resonators are provided in Fig. \ref{Fig1}a and \ref{Fig1}d insets. For the coupled 2-ring device, the round trip length of ring A is 9.5 mm, which corresponds to a free spectral range (FSR) of $\sim$20 GHz, and ring B is 0.5\% longer than ring A. For the 3-ring device, the rightmost ring has a circumference of 9.5 mm, and each other ring is 0.3\% longer than its right neighbor. The rings feature high intrinsic $Q$ factors exceeding 75 million, but individually each ring does not support bright soliton formation around 1550 nm due to the strong normal dispersion associated with the low confinement waveguide structure (see Methods and Extended Data Fig. \ref{ExFig1}). Prior studies on similar single-ring structures have generated only dark pulse comb spectra \cite{jin2021hertz}.

The measured resonance frequency dispersion (green points) for the 2-ring system with comparison to theory (solid lines) is shown in Fig. \ref{Fig2}b. The dispersion of the 3-ring resonator is discussed in section II of the Supplementary Information. The measurement is performed using a radio-frequency calibrated interferometer in combination with a wavelength-tunable laser \cite{yi2015soliton}. The coupled resonators produce the two bands measured in Fig. \ref{Fig2}b where three anomalous dispersion windows are highlighted. At each window, soliton steps are observed when scanning the laser frequency over a cavity resonance (see insets in Fig. \ref{Fig2}b). Operation at the longest and shortest wavelength windows (1584.5 nm and 1525.5 nm) was challenging due to low laboratory laser power and as a result, the time duration of the soliton steps for these wavelengths is relatively shorter.

Analysis shows that the average frequency of the two bands (i.e., $\omega_{\mu} \equiv (\omega_{\mu,+}+\omega_{\mu,-})/2$) is given by the mode frequency for a length-averaged resonator at the same mode number (see Supplementary Information, section I). We note that averaging the frequencies of the two bands removes the effect of the coupling entirely, and the resulting average dispersion shown in Fig. \ref{Fig2}b (orange points) closely matches a parabolic-shaped dispersion curve (orange curve). Accordingly, this average frequency can be described by a second-order dispersion model:
\begin{equation}
\omega_\mu \approx \omega_0 + D_1 \mu + \frac{1}{2} D_2 \mu^2
\label{dispersion_avg}
\end{equation}
where $\omega_0$ is the mode frequency at $\mu=0$ and $\mu$ is a relative mode number referenced to the frequency degeneracy at 1552.3 nm. $D_1$ is the length-averaged FSR for the resonator at $\mu = 0$, $D_2 = -c D_1^2 \beta_2/n_\mathrm{g}$ is the second-order dispersion parameter at $\mu = 0$ with group velocity dispersion $\beta_2$ and waveguide group index $n_\mathrm{g}$. 

On the other hand, the effect of the coupling is made clearer by plotting the mode frequencies relative to the averaged frequency (i.e., relative mode frequency $\omega_{\mu,\pm} - \omega_{\mu}$) as shown in Fig. \ref{Fig2}c. Without inter-ring coupling, the relative mode frequencies of the single rings appear as straight lines on the mode spectrum plot. These lines are related to the straight dashed lines for the uncoupled resonators in Fig. \ref{Fig2}a. Their positive and negative slopes in Fig. \ref{Fig2}c result from removing a linear component of dispersion in this plot given by the average FSR, $D_1$. Mode number walk-off also causes the lines to vertically wrap around at $\pm D_1/2$. For the rings used here, the length of ring B is $0.5\%$ longer than ring A, and frequency degeneracy of the rings occurs every 200 ring A modes (or every 201 ring B modes). The introduction of coupling opens gaps at all frequency degeneracies, regardless of whether the absolute mode number is matched. 

More detailed analysis shows that each of the gap widths equals $2G \equiv g_\mathrm{co}L_\mathrm{co}D_1/\pi$, where $g_\mathrm{co}$ is the coupling strength per unit length and $L_\mathrm{co}$ is the effective coupler length.
The full dispersion relation is found to be (see Supplementary Information, section I):
\begin{equation}
\omega_{\mu,\pm} = \omega_\mu \pm \frac{D_1}{2\pi}\arccos\left[\cos(g_\mathrm{co}L_\mathrm{co})\cos\left(2\pi \epsilon\mu\right)\right]
\label{dispersion_gap}
\end{equation}
where $\epsilon = (L_\mathrm{B} - L_\mathrm{A})/(L_\mathrm{B} + L_\mathrm{A})$ is the length contrast of the rings, and $L_\mathrm{A}$ ($L_\mathrm{B}$) is the length of ring A (B). For the current design $\epsilon = 1/401$, and the gap is modulated with respect to mode number with period $\epsilon^{-1} = 401$ (corresponding to 8 THz in the spectrum). The small length contrast $\epsilon$ guarantees the wide spectral range of the anomalous dispersion window. Overall, there is very good agreement between the model and the measured data in Fig. \ref{Fig2}b and Fig. \ref{Fig2}c, and the fitting allows determination of key resonator parameters (see figure caption).
As an aside, the spectral gap is smaller at larger mode numbers, which can be attributed to the wavelength dependence of $g_\mathrm{co}$, as shorter wavelength results in stronger mode confinement, and hence smaller coupling with the adjacent waveguide. When combined with the original normal dispersion of each ring, the net dispersion for coupled system remains anomalous around $\mu$ = $0$ and $400$ for the upper branch and around $\mu$ = $-200$ and $200$ for the lower branch.

Besides the observation of soliton steps (Fig. \ref{Fig2}b), microcomb spectra measured around 1550 nm for through port (ring A) and drop port (ring B) are presented in Fig. \ref{Fig1}b. The detailed experimental setup is provided in Extended Data Fig. \ref{ExFig3}. The microcomb was stabilized by measuring comb power from the through port and feeding back to the pump laser frequency, which controls the pump-cavity offset frequency \cite{yi2016active}. The theoretical pulse width of the comb spectra in the figure is estimated to be $\sim$ 250 fs. The radio frequency spectrum of the soliton beatnote is presented in Fig. \ref{Fig1}c, indicating good coherence. Microcomb spectra measured at other pump-cavity offset frequencies, and using another device are presented in the Extended Data Fig. \ref{ExFig4}. 

Through port and drop port spectra correspond to pulses in ring A and ring B, and show these pulses are both different from each other and deviate from the conventional sech$^2$ shape of Kerr solitons. The through port spectrum is stronger (weaker) than the drop port at shorter (longer) wavelengths. This is a result of this system representing a new version of the Dirac soliton \cite{wang2020dirac} as discussed in the Methods section.

In Fig.\ref{Fig1}b, two strong dispersive waves (DWs) are observed near 1526 nm and 1577 nm. These correspond to spectral locations where modes of the coupled resonator phase-match to the soliton comb line. For comparison, the dispersion in the vicinity of the comb spectrum has been overlaid in the figure. The DWs broaden the soliton spectrum and provide higher power comb lines (1.5 $\mu$W on-chip power at shorter wavelength and 5.4 $\mu$W at longer wavelength), which is advantageous for application to optical frequency division \cite{diddams2020optical}. To further confirm coherence, the radio frequency spectrum of the soliton beatnote is presented (Fig. \ref{Fig1}c). Finally, the soliton $\mathcal{S}$-resonance and $\mathcal{C}$-resonance \cite{Kippenberg_NP2017universal} were measured using a vector network analyzer. Plots of their relative frequencies versus laser-cavity detuning are given in Extended Data Fig. \ref{Fig_VNA}.

Comb generation in the 3-ring system was also demonstrated (see Fig. \ref{Fig1}d). Here, the coupling on both sides of the middle ring creates local anomalous dispersion windows (Supplementary Information, section II). Fig. \ref{Fig1}e shows the spectrum of three pulses as measured from the center ring. The measured dispersion is also included in the figure. The pump laser wavelength is several nanometers away from the anomalous dispersion center frequency, and, as a result, the spectrum features only one dispersive wave at the shorter wavelength side. The radio frequency spectrum of the soliton beatnote is presented in Fig. \ref{Fig1}f, indicating good coherence.

\section{Pulse pairs and multi-partite states}

This section describes a time domain picture of the coupled-ring system. Besides providing a complementary physical picture (to the dispersion analysis above), simulations of mode locking show microcombs form as phase-locked pulse pairs where the pulses have opposite phases. The pair viewpoint provides a powerful framework for visualization of mode locking that readily explains observable multi pulse-pair states and higher dimensional systems comprising multiple coupled cavities. 

Simulations of pulse propagation in the 2-ring system are presented in Fig. \ref{Fig3}a. Here, the ring FSRs and couplings are those of the experimental system studied in Fig. \ref{Fig2} b,c, and excitation occurs for the mode $\mu =0$. As shown in Fig. \ref{Fig3}b, each pulse undergoes shape, chirp, and pulse width variations that repeat upon each round trip. Before entering the coupling region (point I in Fig. \ref{Fig3}a), the chirp of both pulses has increased due to uncompensated Kerr nonlinearity from propagation in normal dispersion waveguides of each ring. Pulse chirp is indicated in the lower panel of Fig. \ref{Fig3}a, where the color represents instantaneous frequency. The pulse in ring B (red) also lags behind its counterpart in ring A (blue) due to the difference in ring lengths. However, upon entering the coupling region (point II), the ring B (A) pulse accelerates (decelerates) and becomes the leading (lagging) pulse when exiting the coupling region (point III). In the meantime, the chirp of both pulses decreases through the coupling region. Upon exiting the coupling region, the pulses propagate in their respective waveguides (point IV) where chirp increases as the pulses circle back through point I. 
Detailed numerical simulations (see Methods) are used to further explore and confirm the pulse pair evolution (Fig. \ref{Fig3}b).

This picture of pairwise round trip compensation of normal dispersion enables understanding of how compensation works for multi-pair systems as well as for higher dimensions with additional ring cavities. Specifically, it constrains the ways these states are allowed to form. For example, consider the coupled-ring states in Fig. \ref{Fig4}a,b wherein 2 pulse pairs circulate in a mirror-image like fashion to form the observed spectra. Here, to reduce confusion with corresponding multi-pulse soliton systems, we adopt the nomenclature that a single pulse pair in a 2 ring system is a bipartite single soliton (see Fig. \ref{Fig1}a,b), while multi-pair states in the same are bipartite multi soliton systems. Specfically, the states in Fig. \ref{Fig4}a,b are bipartite 2 soliton states. The state in Fig. \ref{Fig4}b is moreover a bipartite 2-soliton crystal. Notice that the requirements imposed on pulse pairing allow a one-to-one correspondence between conventional multi-soliton states and bipartite states, since the pulse configurations in each ring resonator must mirror image its neighboring ring. 

The same is true for higher dimensional systems. 
For example, three pulses compensate normal dispersion by alternating their pairwise coupling as illustrated in Fig. \ref{Fig1}d. Here, the outer ring pulses experience compensation once per cycle, but the inner ring pulse experiences compensation twice per cycle. Moreover, the pairwise compensation works when additional pulses are added to each cavity. For example, measurement of tripartite 2 soliton, 3 soliton and a 4 soliton crystal state (containing respectively 6, 9 and 12 pulses) are presented in Fig. \ref{Fig4}c-f. Notice that the measured comb line spacing (79.93 GHz) for the crystal state is four times the FSR of a single ring as is consistent with a conventional 4 soliton crystal state.

\section{Discussion}

In summary, we have observed a new type of microcavity soliton that mode locks as pulse pairs distributed spatially over multiple ring resonators. The requirement to compensate overall normal dispersion of the rings requires that the pulses in each ring arrange themselves as a mirror image of the pulses in neighboring rings.
Partial coupling of the resonators creates a situation in which ring resonator mode number is not conserved  and this enables recurring spectral windows where the pairs can be formed. The presented bright soliton results use the CMOS-ready process that has previously been restricted to only dark pulse generation. The ability to distribute coherent pulses over multiple rings with individual taps and with simultaneous pulse formation at multiple wavelengths presents new opportunities for soliton science and microcomb applications. 

\bibliography{main.bib}

\newpage

\noindent\textbf{Methods}

\noindent \textbf{Resonator design.} The rings consist of Si$_3$N$_4$ waveguides (2800 nm width and 100 nm thickness) embedded in silica and formed into a racetrack shape. 
The waveguide cross-section only supports one polarization mode.
Detailed information on fabrication steps can be found elsewhere \cite{jin2021hertz}. For the two-ring device, ring A has a circumference of 9.5 mm, and ring B is 0.5$\%$ longer. For the three-ring device, the rightmost ring has a circumference of 9.5 mm, and each other ring is 0.3\% longer than its right neighbor. The adiabatic waveguide bend has the shape of a fifth-degree spline such that the curvature is continuous along the curve and transition loss is minimized. The gap between the inner edges of the two waveguides in the coupling region is 2400 nm, and the effective coupling length is 1.0 mm including
contributions from the adiabatic bend (which is $10.5\%$ of the shortest ring circumference).

The simulated dispersion of straight Si$_3$N$_4$ waveguides with 2.8 $\mu$m width are shown in Extended Data Fig. \ref{ExFig1}a.
For these calculations, the effective index of the fundamental TE mode was calculated and the group velocity dispersion determined through $\beta_2 = \lambda^3/(2\pi c^2) \partial^2 n_\mathrm{wg}/\partial \lambda^2$, where $\lambda$ is the vacuum wavelength. For waveguides with thickness under 780 nm, the fundamental TE mode always features normal dispersion in the C-band. To maintain high optical $Q$ factors, the waveguide thickness is about 100 nm for the current process, which places the waveguide deep into the normal dispersion region.

Simulations of the waveguide coupling rate $g_\mathrm{co}$ with 2.4 $\mu$m coupling gap are presented in Extended Data Fig. \ref{ExFig1}b.
The effective index of the two supermodes at the coupling region is calculated, and the coupling rate $g_\mathrm{co}$ is related to the index difference of the supermodes $\Delta n_\mathrm{wg}$ by $g_\mathrm{co} = \Delta n_\mathrm{wg} \pi/\lambda$.
With a thinner waveguide or a longer wavelength, the optical confinement is weaker, leading to a larger coupling strength and larger spectral gap width.

\medskip

\noindent \textbf{Dispersion measurement and fitting.} The dispersion is measured by sweeping a mode-hop-free laser while pumping the resonator, recording the mode positions from the transmission signal, and comparing it against a calibrated Mach-Zehnder interferometer \cite{yi2015soliton}.
The averaged mode frequencies are fitted by a second-order dispersion model given by Eq. (\ref{dispersion_avg}) with $D_1 = 2\pi\times 19.9766$ GHz and $D_2 = 2\pi\times(-283.0)$ kHz. The relative frequencies are fitted with Eq. (\ref{dispersion_gap}), where we assume that the coupling is exponentially decaying with respect to mode number:
\begin{equation}
g_\mathrm{co} = g_\mathrm{co,0}\exp(-\mu/\mu_g)
\end{equation}
where $\mu_g$ gives a decay scale. The fitting uses $g_\mathrm{co,0}$, $\mu_g$ and the crossing center position as fitting parameters, while $D_1$ and $D_2$ are derived from the mode frequency average fitting and $\epsilon = 1/401$ is taken from design values. Fitting gives $g_\mathrm{co,0}L_\mathrm{co} = 0.954$ and $\mu_g = 1196$. The coupling is equivalent to a $33\%:67\%$ coupler near $\mu = 0$, and the coupling rate increases by $5.4\%$ for every $10$ nm increased near $1550$ nm. The coupling rate and decaying scale are close to simulation results ($g_\mathrm{co,0}L_\mathrm{co} = 0.782$, $5.5\%$ increase per $10$ nm; see Extended Data Fig. \ref{ExFig1}b). Differences between measured and simulated values may result from refractive index and layer thickness variations.

\noindent \textbf{Dynamics of the soliton pulse pair.} The optical fields in the two rings are governed by the coupled nonlinear wave equations:
\begin{align}
    \frac{\partial E_{\rm A}}{\partial t} =& -\left(\frac{\kappa}{2} + i \delta \omega_{\rm A}\right) E_{\rm A} - v_{\rm g} \frac{\partial E_{\rm A}}{\partial z} - i \frac{\beta_2 v_{\rm g}^3}{2} \frac{\partial^2 E_{\rm A} }{\partial z^2} \nonumber \\ &+ i g_{\rm co} v_{\rm g} \chi_{\rm co}(z) E_{\rm B} + i g_{\rm NL} \vert E_{\rm A} \vert^2 E_{\rm A} + F
    \label{LLE_A}\\
    \frac{\partial E_{\rm B}}{\partial t} =& -\left(\frac{\kappa}{2} + i \delta \omega_{\rm B}\right) E_{\rm B} - v_{\rm g} \frac{\partial E_{\rm B}}{\partial z} - i \frac{\beta_2 v_{\rm g}^3}{2} \frac{\partial^2 E_{\rm B} }{\partial z^2} \nonumber \\ &+ i g_{\rm co} v_{\rm g} \chi_{\rm co}(z) E_{\rm A} + i g_{\rm NL} \vert E_{\rm B} \vert^2 E_{\rm B}
    \label{LLE_B}
\end{align}
accompanied by periodic boundary conditions in the $z$ direction, where $E_{\rm A,B}$ denotes the optical field in the two rings normalized to photon numbers in the corresponding length-averaged ring, $\kappa = \kappa_{\rm in} +  \kappa_{\rm ex}$ is the loss rate (sum of intrinsic and external loss) for the individual rings (assumed to be identical for ring A and B), which can be linked to the quality factors via $\kappa=\omega_0/Q$, $\kappa_{\rm in}=\omega_0/Q_{\rm in}$, and $\kappa_{\rm ex}=\omega_0/Q_{\rm ex}$. Also, $\delta \omega_{\rm A,B} = \omega_{0{\rm A,B}} - \omega_{\rm p}$ is the pump laser detuning, $v_{\rm g}=c/n_{\rm g}$ is the group velocity of the waveguide, $z \in [0,L_{\rm A,B})$ is the resonator coordinate with $L_{\rm A,B}$ the ring length, $\beta_2$ is the waveguide group velocity dispersion, $g_{\rm co}$ is the coupling strength between the two waveguides in the coupling region, $\chi_{\rm co}(z)$ is the indicator function with value 1 in the coupling region and 0 elsewhere, $g_{\rm NL} = \hbar \omega_0^2 D_1 n_2/(2\pi n_{\rm g} A_{\rm eff})$ is the nonlinear coefficient with $A_{\rm eff}$ being the effective mode area, and $F = \sqrt{\kappa_{\rm ex} P_{\rm in}/\hbar \omega_0}$ is the pump term where $P_{\rm in}$ is the on-chip pump power.
For simplicity, the pump and loss terms are averaged over the entire resonator without considering the detailed coupling geometry between the rings and the bus waveguides.

To demonstrate that the resulting soliton resembles the optical Dirac soliton \cite{wang2020dirac}, we will convert the above equations into a form that is analogous to the Dirac equation in quantum field theory. We start by defining a common roundtrip variable $\theta$ for both resonators, with $\theta = 2\pi z/L_\mathrm{A}$ for ring A and $\theta = 2\pi z/L_\mathrm{B}$ for ring B. With this change, the LLE reads
\begin{align}
    \frac{\partial E_{\rm A}}{\partial t} =& -\left(\frac{\kappa}{2} + i \delta \omega_{\rm A}\right) E_{\rm A} \nonumber \\
    &- \frac{D_1}{1-\epsilon} \frac{\partial E_{\rm A}}{\partial \theta} + i \frac{D_2}{2(1-\epsilon)^2} \frac{\partial^2 E_{\rm A} }{\partial \theta^2} \nonumber \\
    &+ i g_{\rm co} v_{\rm g} \chi_{\rm co}(\theta) E_{\rm B} + i g_{\rm NL} \vert E_{\rm A} \vert^2 E_{\rm A} + F
\end{align}
and similarly for $E_{\rm B}$ with $\epsilon$ replaced by $-\epsilon$ and pump term dropped. The unified roundtrip variable breaks the correspondence of waveguide sections in the coupling region, but these have been neglected as the pulse width is much larger compared to the ring length difference (Fig. \ref{Fig2}c). Switching to the co-moving frame of the pulse [$\psi_{\rm A,B}(\theta, t) \equiv E_{\rm A,B}(\theta + D_1 t, t)$] leads to
\begin{align}
    \frac{\partial \psi_{\rm A}}{\partial t} \approx& -\left(\frac{\kappa}{2} + i \delta \omega_{\rm A}\right) \psi_{\rm A} - \epsilon D_1 \frac{\partial \psi_{\rm A}}{\partial \theta} + i \frac{D_2}{2} \frac{\partial^2 \psi_{\rm A} }{\partial \theta^2} \nonumber \\
    &+ i G \psi_{\rm B} + i g_{\rm NL} \vert \psi_{\rm A} \vert^2 \psi_{\rm A} + F
\end{align}
and similarly for $E_{\rm B}$, where we retain the lowest order of $\epsilon$ and further assume that the pulse varies slowly within one round trip such that the effect of coupling is averaged over the resonator length (i.e., uniform coupling which conserves the mode number). Finally, shifting the wavevector and frequency reference ($\tilde{\psi}_{\rm A,B} \equiv \psi_{\rm A,B} \exp(ik_0\theta - i\omega_0 t)$) gives
\begin{align}
    \frac{\partial \tilde{\psi}_{\rm A}}{\partial t} \approx& -i\left(\delta \omega_{\rm A} - \epsilon D_1 k_0 + \omega_0\right) \tilde{\psi}_{\rm A} - \epsilon D_1 \frac{\partial \tilde{\psi}_{\rm A}}{\partial \theta} + i G \tilde{\psi}_{\rm B} \nonumber \\
    & + i g_{\rm NL} \vert \tilde{\psi}_{\rm A} \vert^2 \tilde{\psi}_{\rm A} \nonumber \\
    & - \frac{\kappa}{2}\tilde{\psi}_{\rm A} + i \frac{D_2}{2} \frac{\partial^2 \tilde{\psi}_{\rm A} }{\partial \theta^2} + F\exp(ik_0\theta-i\omega_0 t) \\
    \frac{\partial \tilde{\psi}_{\rm B}}{\partial t} \approx& -i\left(\delta \omega_{\rm B} + \epsilon D_1 k_0 + \omega_0\right) \tilde{\psi}_{\rm B} + \epsilon D_1 \frac{\partial \tilde{\psi}_{\rm B}}{\partial \theta} + i G \tilde{\psi}_{\rm A} \nonumber \\
    & + i g_{\rm NL} \vert \tilde{\psi}_{\rm B} \vert^2 \tilde{\psi}_{\rm B} \nonumber \\
    & - \frac{\kappa}{2}\tilde{\psi}_{\rm B} + i \frac{D_2}{2} \frac{\partial^2 \tilde{\psi}_{\rm B} }{\partial \theta^2}
\end{align}
where we assume that we are pumping near the crossing center such that $\epsilon D_1 \ll D_2 k_0$ and high-order terms in $k_0$ could be neglected. Choosing $k_0 = (\delta \omega_{\rm A} - \delta \omega_{\rm B})/(2\epsilon D_1)$ and $\omega_0 = -(\delta \omega_{\rm A} + \delta \omega_{\rm B})/2$ removes the effective detuning terms from the two equations.

\begin{figure}[t!]
\begin{centering}
\includegraphics[width=\linewidth]{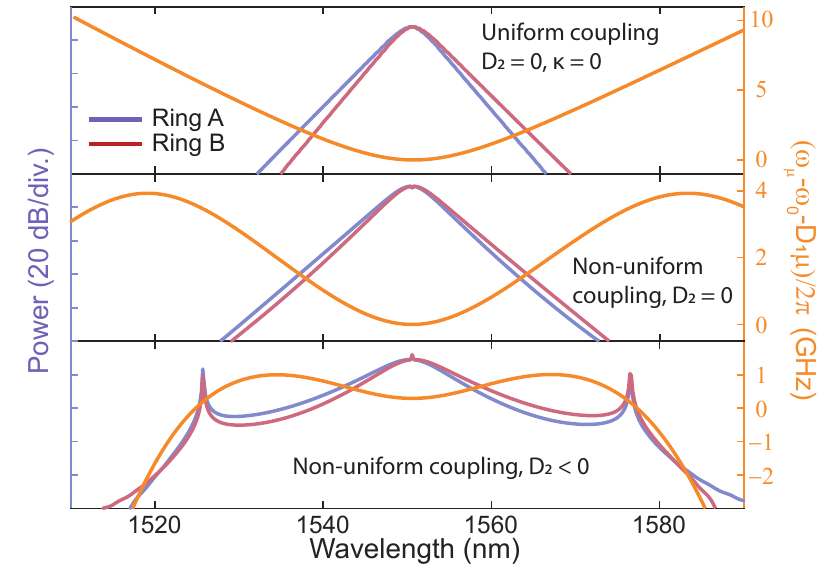}
\captionsetup{singlelinecheck=no, justification = RaggedRight}
\caption{{\bf Simulated optical spectra and dispersion relation for Dirac solitons assuming different levels of approximations in the model.} 
Top panel: Uniform coupling between two rings (mode number conservation), without pump and loss, and with zero second-order dispersion.
Middle panel: Non-uniform coupling between two rings (mode number non-conservation), with pump and loss included, and with zero second-order dispersion. Recurring dispersion relations can be observed but the spectrum is free of strong dispersive waves.
Bottom panel: Non-uniform coupling between two rings (mode number non-conservation), with pump and loss, and with negative second-order dispersion [i.e. full Eqs. (\ref{LLE_A}) and (\ref{LLE_B})].
}
\label{Fig_Dirac}
\end{centering}
\end{figure}

This can now be compared to the massive Dirac equation in $1+1$ dimension written in a chiral basis \cite{de1986field}:
\begin{align}
    \partial_t \psi_{\rm L} = -c\partial_x \psi_{\rm L} + i\frac{Mc^2}{\hbar}\psi_{\rm R} \\
    \partial_t \psi_{\rm R} = +c\partial_x \psi_{\rm R} + i\frac{Mc^2}{\hbar}\psi_{\rm L}
\end{align}
where $M$ is interpreted as the mass, and corresponds to the coupling term (the massless Dirac equation with $M=0$ would correspond to an uncoupled system with frequency gap closed). The momentum term corresponds to the FSR difference. The nonlinear term converts the equation into a nonlinear Dirac equation, although there is no exact analogue of the self-phase modulation in quantum field theory as this contradicts the Pauli exclusion principle. Loss, pump and second-order dispersion terms do not have analogues in the nonlinear Dirac equation, and could be treated as perturbations for the soliton dynamics. For example, $D_2$ is no longer the dominant contribution to dispersion near the mode crossing center. We note that these terms do not change the qualitative features of the generated soliton, therefore establishing the link between the current soliton and the optical Dirac soliton previously studied \cite{wang2020dirac}. A comparison of the simulated soliton spectral profile using different levels of approximation can be found in Fig. \ref{Fig_Dirac}.

\medskip

\noindent \textbf{Simulations of soliton pulse pair formation.} Numerical simulations have been performed based on the nonlinear wave equations [Eqs. (\ref{LLE_A}) and (\ref{LLE_B})] and the results are used for plotting Fig. \ref{Fig2}c. For simplicity, the coupling is assumed to be wavelength independent ($g_{\rm co}=g_{\rm co,0}$), which makes understanding the dispersion compensation in the coupling region more transparent. Parameters used for numerical simulations are: $\omega_0=2\pi\times193.34$ THz, $Q_{\rm in}=75$ M, $Q_{\rm ex}=45$ M, $\delta \omega_{\rm A}=\delta \omega_{\rm B}=12.5\kappa - G$ where $G$ is the half gap created by the coupling (pump is red-detuned with respect to the upper branch resonance by $12.5\kappa$), $D_1=2\pi\times19.9766$ GHz, $D_2=-2\pi\times283.0$ kHz, $n_{\rm g}=1.575$, $P_{\rm in} = 300$ mW, $g_{\rm NL}=0.0277$ ${\rm s}^{-1}$, and $g_{\rm co,0}=0.954$ ${\rm mm}^{-1}$.

\medskip

\noindent \textbf{Soliton dynamics in the coupling region.} In the coupling region where linear interaction is dominant in the soliton dynamics, the coupled LLE can be reduced to:
\begin{align}
    \frac{\partial E_{\rm A}}{\partial z} + \frac{1}{v_{\rm g}} \frac{\partial E_{\rm A}}{\partial t} & =  i g_{\rm co} E_{\rm B}
    \\
    \frac{\partial E_{\rm B}}{\partial z} + \frac{1}{v_{\rm g}} \frac{\partial E_{\rm B}}{\partial t} & =  i g_{\rm co} E_{\rm A}
\end{align}
where $z=0$ denotes the beginning of the coupling region. Note that $g_\mathrm{co}$ here is assumed to be wavelength independent for simplicity. The optical fields at $z$ can be related to the incident fields ($z=0$) as
\begin{align} \label{couple_region_theory_A}
    E_{\rm A}(z,t)&=\cos{(g_{\rm co}z)}E_{\rm A}(0,t') + i \sin{(g_{\rm co}z)}E_{\rm B}(0,t') \\
    \label{couple_region_theory_B}
    E_{\rm B}(z,t)&=\cos{(g_{\rm co}z)}E_{\rm B}(0,t') + i \sin{(g_{\rm co}z)}E_{\rm A}(0,t')
\end{align}
where $t'=t-{z}/{v_{\rm g}}$ is the retarded time. The evolution of soliton properties with propagation distance plotted in Fig. \ref{Fig2}c is obtained from Eqs. (\ref{couple_region_theory_A}) and (\ref{couple_region_theory_B}), with initial conditions $E_{\rm A,B}(0,t')$ taken from simulations, and shows good agreement with the simulation results using Eq. (\ref{LLE_A}) and (\ref{LLE_B}).

\medskip

\vspace{3 mm}

\noindent \textbf{Data Availability}
The data that supports the plots within this paper and other findings of this study are available from the corresponding author upon reasonable request.

\vspace{1 mm}

\noindent \textbf{Code Availability}
The codes that support findings of this study are available from the corresponding author upon reasonable request.

\vspace{1 mm}
\noindent \textbf{Acknowledgments}
The authors thank Chao Xiang for providing the optical images of the resonators.
This work is supported by the Defense Advanced Research Projects Agency (HR0011-22-2-0009), the Defense Threat Reduction Agency-Joint Science and Technology Office for Chemical and Biological Defense (grant No. HD-TRA11810047), the Air Force Office of Scientific Research (FA9550-18-1-0353) and the Kavli Nanoscience Institute at Caltech.
The content of the information does not necessarily reflect the position or the policy of the federal government, and no official endorsement should be inferred.
\vspace{1 mm}

\noindent\textbf{Author Contributions} Concepts were developed by Z.Y., M.G., Y.Y., H.W., W.J., J.B. and K.V. Measurements and modeling were performed by Z.Y., M.G., Y.Y., H.W., W.J., Q.-X.J. Structures were designed by W.J. and H.W. Sample preparation and logistical support provided by A.F. and M.P. All authors contributed to the writing of the manuscript. The project was supervised by J.B. and K.V. 
\vspace{1 mm}

\noindent \textbf{Competing Interests} The authors declare no competing interests.

\vspace{1 mm}

\noindent \textbf{Author Information} Correspondence and requests for materials should be addressed to KV (vahala@caltech.edu).

\clearpage
\onecolumngrid
\renewcommand{\figurename}{\bf Extended Data Fig.}
\setcounter{figure}{0}

\newpage

\begin{figure*}[t!]
\begin{centering}
\includegraphics[width=170mm]{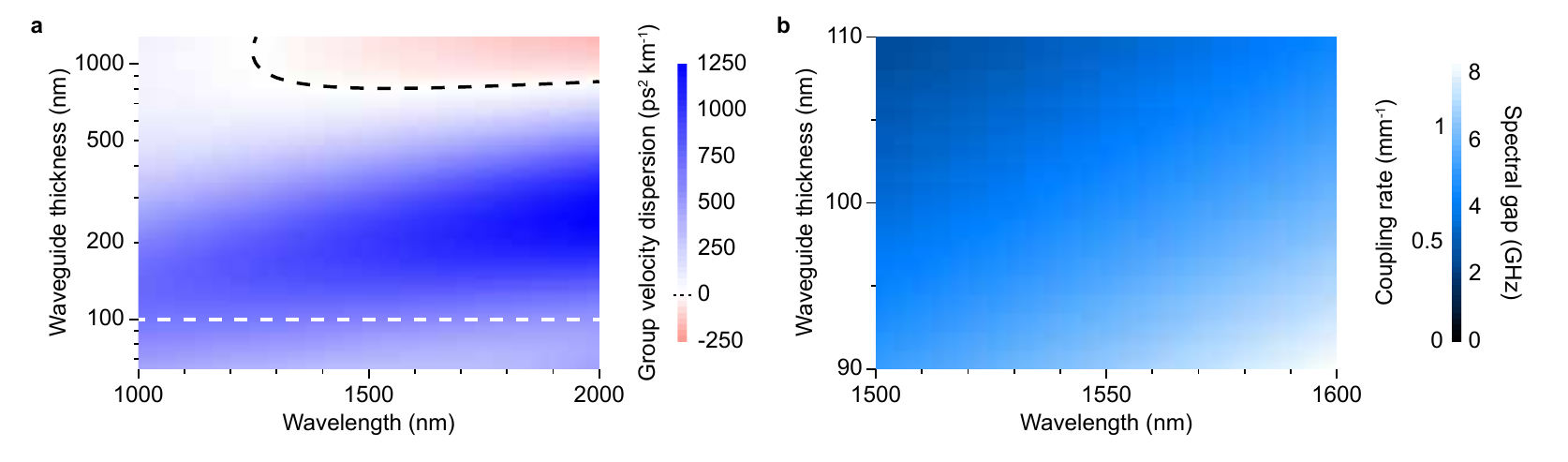}
\captionsetup{singlelinecheck=off, justification = RaggedRight}
\caption{{\bf Dispersion and coupling characteristics of the ring waveguide.}
\textbf{a}  
Finite element simulation results for dispersions of straight Si$_3$N$_4$ waveguides with fixed width (2.8 $\mu$m) as a function of wavelength and waveguide thickness.
The zero-dispersion boundary is marked as the black dashed curve.
Nominal waveguide thickness (100 nm) for the current process is marked as the white dashed line.
\textbf{b} Numerical simulations of the waveguide coupling rate $g_\mathrm{co}$ and the corresponding spectral gap ($2G = g_\mathrm{co}L_\mathrm{co}D_1/\pi$, with $L_\mathrm{co} = 1.0$ mm and $D_1 = 2\pi \times 20$ GHz) are plotted as a function of wavelength and waveguide thickness. The gap between waveguides is 2.4 $\mu$m.
}
\label{ExFig1}
\end{centering}
\end{figure*}

\clearpage
\newpage

\begin{figure*}[t!]
\begin{centering}
\includegraphics[width=170mm]{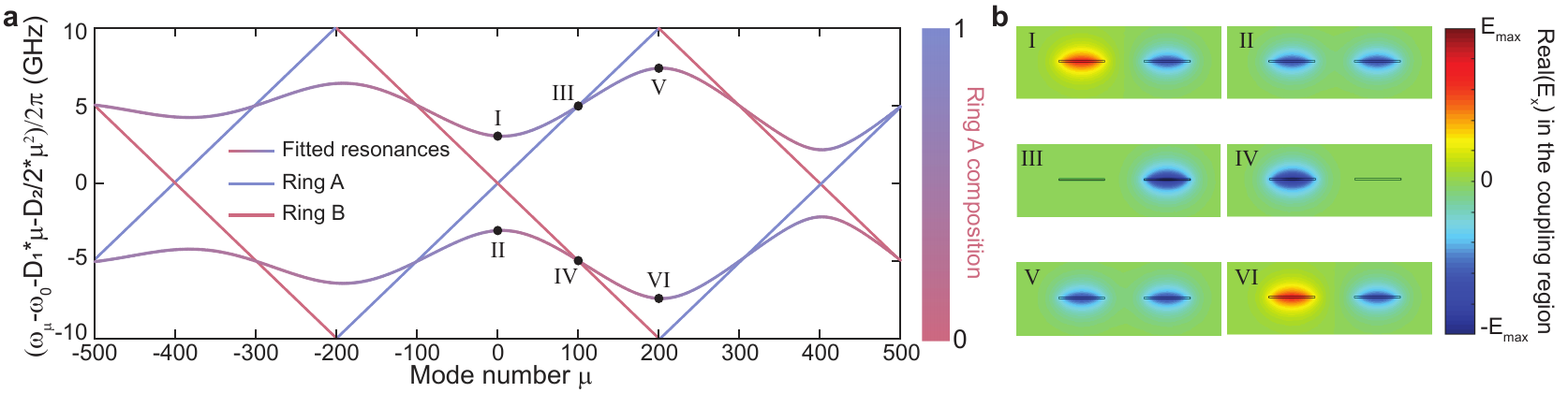}
\captionsetup{singlelinecheck=off, justification = RaggedRight}
\caption{{\bf Illustration of mode hybridization in the coupling region.}
\textbf{a} Fitted optical resonance frequency dispersion of the coupled resonator (solid curves) and fitted mode frequency dispersion of the single rings (red and blue lines) plotted versus relative mode number $\mu$. 
These plots are the same as Fig. \ref{Fig2}c in the main text.
\textbf{b} Cross-sectional view of simulated electric field amplitudes in the coupled region at mode numbers indicated in panel \textbf{a} by the black points. The right (left) waveguide belongs to ring A (B). The waveguide geometry used here is 2.8$\mu$m*0.1$\mu$m, and the gap between two waveguides is 2.4 $\mu$m. At the crossing center (I, II, V and VI), two waveguides have the same field intensity and the opposite (same) phase. When hybrid mode frequencies meet the single-ring resonances (III and IV), the electrical field at the coupled region is contributed by a single ring.
}
\label{ExFig2}
\end{centering}
\end{figure*}

\clearpage
\newpage

\begin{figure*}[t!]
\begin{centering}
\includegraphics[width=170mm]{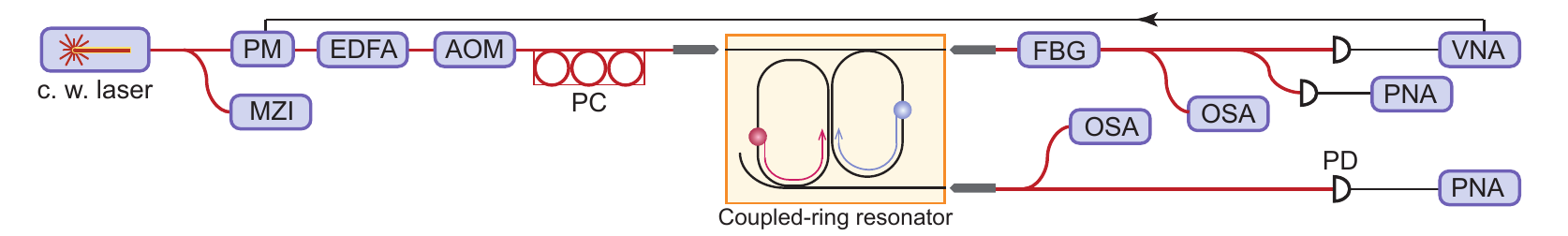}
\captionsetup{singlelinecheck=off, justification = RaggedRight}
\caption{{\bf Experiment setup for generation and characterization of the bright solitons in the coupled-ring resonators.}
The output of a continuous wave fiber laser is amplified by an erbium-doped fiber amplifier (EDFA), and then coupled to the input of the bus waveguide. 
Soliton power was collected by a lensed fiber from the through port and drop port.
The through-port output is filtered by a fiber Bragg grating (FBG) to isolate comb power from the pump. 
The comb power from through port as well as output from drop port are sent to a phase noise analyzer (PNA), optical spectrum analyzer (OSA) and vector network analyzer (VNA) to characterize beat note phase noise, comb spectrum and soliton resonances, respectively. The VNA-controlled frequency modulation to the pump laser is applied through a phase modulator (PM), and the measured responses are presented in Fig. \ref{Fig3}a.
MZI: Mach–Zehnder interferometer, AOM: acousto-optical modulator, PD: photodetector.
}
\label{ExFig3}
\end{centering}
\end{figure*}

\clearpage
\newpage

\begin{figure*}[t!]
\begin{centering}
\includegraphics[width=170mm]{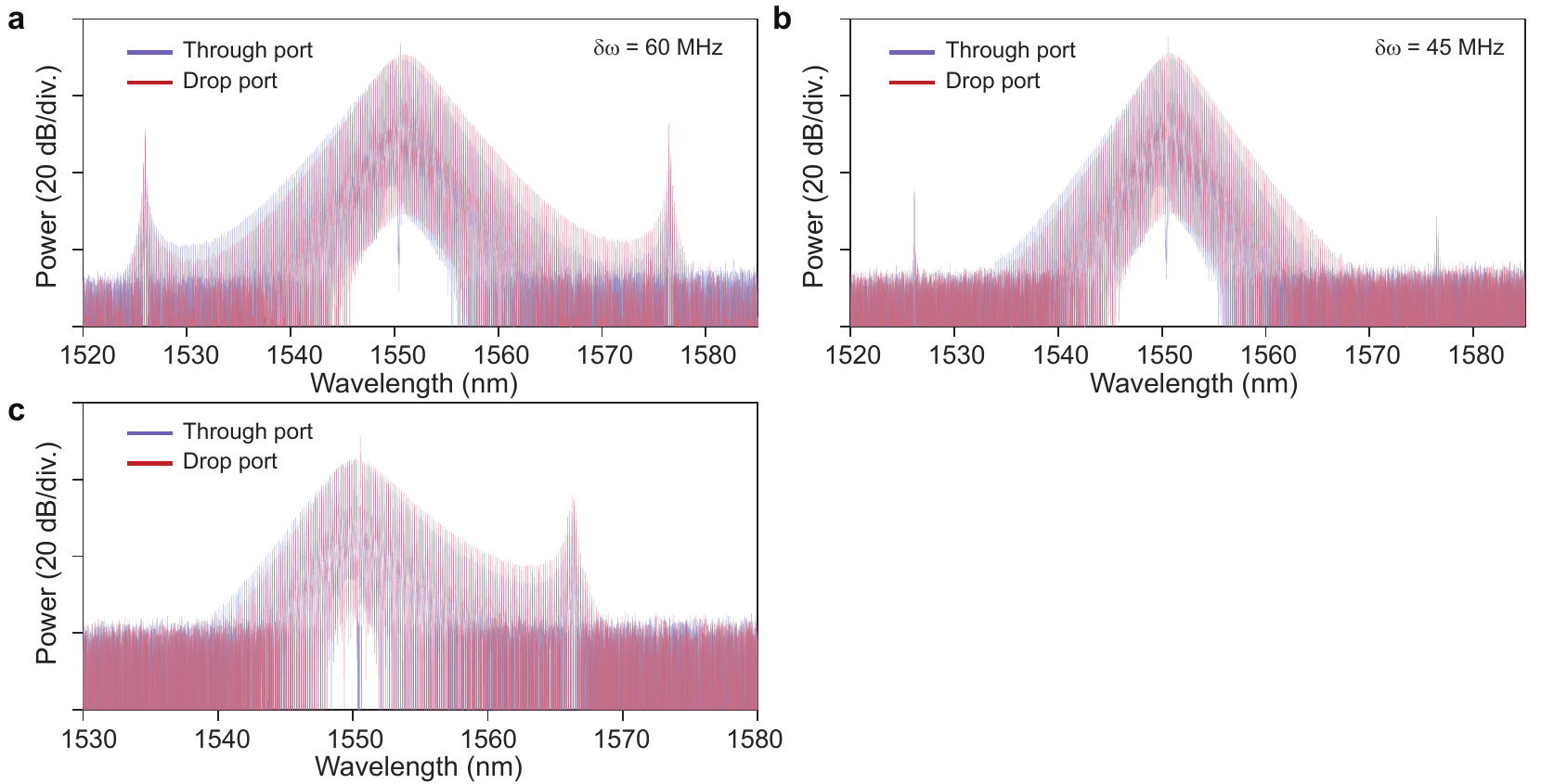}
\captionsetup{singlelinecheck=off, justification = RaggedRight}
\caption{{\bf Additional optical spectra of the solitons in the two-ring coupled resonator.}
\textbf{a, b} Soliton pulse pair optical spectra in two-ring coupled resonator at different pump laser detunings ($\delta \omega$), for comparison to the optical spectrum in Fig. \ref{Fig1}b in the main text (where $\delta\omega= $75 MHz).
\textbf{c} Soliton pulse pair optical spectra generated in another device wherein the coupling center wavelength is several nanometers away from the pump laser wavelength. As a result, the spectra feature only one dispersive wave on the longer wavelength side.
}
\label{ExFig4}
\end{centering}
\end{figure*}

\clearpage
\newpage

\begin{figure*}[t!]
\begin{centering}
\includegraphics[width=170mm]{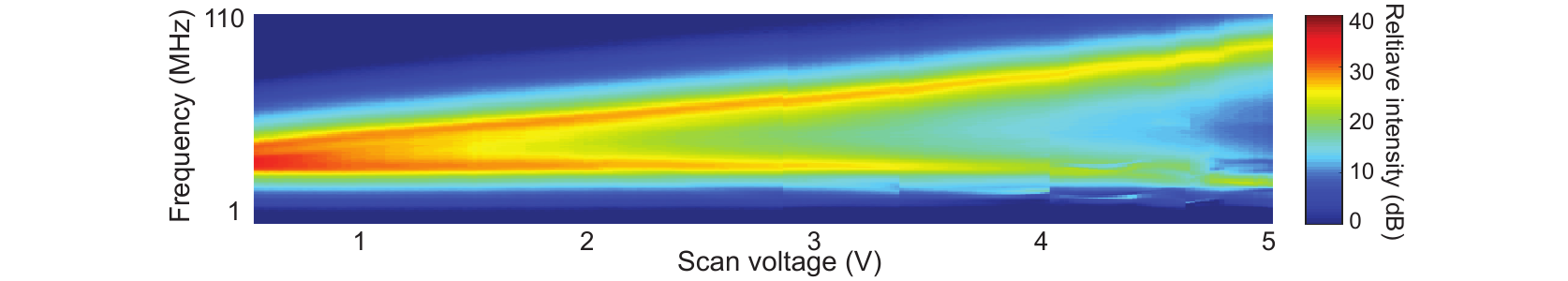}
\captionsetup{singlelinecheck=off, justification = RaggedRight}
\caption{{\bf Characterization of soliton $\mathcal{C}$ and $\mathcal{S}$ resonances at multiple laser-cavity detunings in the 2 ring system for a single pulse pair.}
The relative frequency of the $\mathcal{C}$ and $\mathcal{S}$ resonances are measured using a vector network analyzer and plotted versus tuning voltage.
}
\label{Fig_VNA}
\end{centering}
\end{figure*}

\clearpage
\onecolumngrid
\appendix
\renewcommand{\theequation}{S\arabic{equation}}
\renewcommand{\thefigure}{S\arabic{figure}}
\setcounter{figure}{0}
\setcounter{equation}{0}

\section*{Supplementary Information}

\section{Eigenmodes of a two-ring coupled resonator}

In this section we study the coupling between the two coupled rings and analyze the mode frequencies of the compound system. Eigenfrequencies of coupled resonators have traditionally been calculated from a coupled-mode perspective, where modal coupling are calculated as off-diagonal matrix elements. However, this approach becomes unfeasible in the current system as one longitudinal mode will couple to many modes from the opposite ring because mode number matching is not required. The dependence of coupling with respect to wavelength is also difficult to implement. To circumvent these problems, we instead use a transfer function formalism to determine the mode frequencies.

\begin{figure}[b]
\centering
\includegraphics[width=170mm]{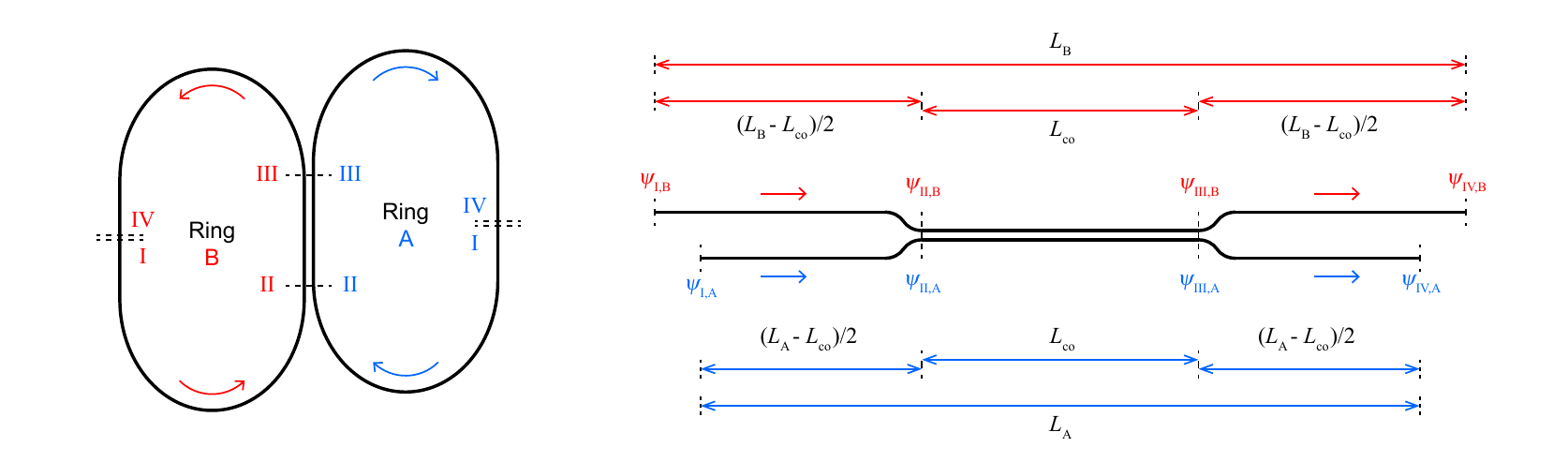}
\captionsetup{singlelinecheck=off, justification = RaggedRight}
\caption{Schematic of the two-ring coupled resonator.
Left panel: Top view of the coupled resonator with key points marked.
Right panel: Schematic of the resonator with straightened waveguides (not to scale). Segment lengths and field amplitudes have been marked.}
\label{FigS1}
\end{figure}

The resonator schematic is shown in Fig. \ref{FigS1}. $L_\mathrm{A}$ and $L_\mathrm{B}$ are the circumferences of the right and left ring, respectively, $L_\mathrm{co}$ is the length of the coupling region, $c$ is the speed of light in vacuum, and $n_\mathrm{wg}(\omega)$ is the effective phase index of the fundamental mode in the SiN waveguide at optical angular frequency $\omega$. By defining a single index along the waveguide, we have neglected the geometric dispersions resulting from bending the waveguide. These have been shown to be small compared to the geometric dispersion induced by waveguide confinement. We now assume that light with a single frequency is propagating in the system. At the points opposite to the coupling region, the field amplitude in each ring is denoted as $\psi_\mathrm{I,A}$ and $\psi_\mathrm{I,B}$. These amplitudes can be assembled into a vector as $\psi_\mathrm{I} = (\psi_\mathrm{I,A}, \psi_\mathrm{I,B})^\mathbf{T}$, where $^\mathbf{T}$ denotes the transpose of a vector or matrix. Similarly, the field just before the coupling part can be found as
\begin{equation}
\begin{pmatrix} \psi_\mathrm{II,A} \\ \psi_\mathrm{II,B} \end{pmatrix}
=
\begin{pmatrix}
e^{in_\mathrm{wg}\omega(L_\mathrm{A}-L_\mathrm{co})/(2c)} & 0 \\
0 & e^{in_\mathrm{wg}\omega(L_\mathrm{B}-L_\mathrm{co})/(2c)}
\end{pmatrix}
\begin{pmatrix} \psi_\mathrm{I,A} \\ \psi_\mathrm{I,B} \end{pmatrix}
\end{equation}
For the coupling section, we denote the coupling rate per unit length as $g_\mathrm{co}$. The coupling depends on $\omega$, and is assumed to be uniform along the coupling section (i.e., boundary effects from adiabatic bends are included in the effective coupling length). The field after the coupling section can be expressed with a matrix exponential:
\begin{equation}
\begin{pmatrix} \psi_\mathrm{III,A} \\ \psi_\mathrm{III,B} \end{pmatrix}
=
\exp{\left[i L_\mathrm{co}
\begin{pmatrix}
n_\mathrm{wg}\omega/c & g_\mathrm{co} \\
g_\mathrm{co} & n_\mathrm{wg}\omega/c
\end{pmatrix}
\right]}
\begin{pmatrix} \psi_\mathrm{II,A} \\ \psi_\mathrm{II,B} \end{pmatrix}
\end{equation}
Finally, returning to the points opposite to the coupling region, the field reads
\begin{equation}
\begin{pmatrix} \psi_\mathrm{IV,A} \\ \psi_\mathrm{IV,B} \end{pmatrix}
=
\begin{pmatrix}
e^{in_\mathrm{wg}\omega(L_\mathrm{A}-L_\mathrm{co})/(2c)} & 0 \\
0 & e^{in_\mathrm{wg}\omega(L_\mathrm{B}-L_\mathrm{co})/(2c)}
\end{pmatrix}
\begin{pmatrix} \psi_\mathrm{III,A} \\ \psi_\mathrm{III,B} \end{pmatrix}
\end{equation}

For modes in the system, we require the state to reproduce itself after one round trip:
\begin{equation}
\psi_\mathrm{IV} = e^{i\Theta} \psi_\mathrm{I}
\end{equation}
This requires finding the eigenvalues of the roundtrip transfer matrix $T$, which is the product of the previous three transfer matrices:
\begin{equation}
\psi_\mathrm{IV} = T\psi_\mathrm{I},\ \ T=e^{in_\mathrm{wg}\omega\overline{L}/c}
\begin{pmatrix}
e^{in_\mathrm{wg}\omega \Delta L/c}\cos(g_\mathrm{co}L_\mathrm{co}) & i\sin(g_\mathrm{co}L_\mathrm{co}) \\
i\sin(g_\mathrm{co}L_\mathrm{co}) & e^{-in_\mathrm{wg}\omega \Delta L/c}\cos(g_\mathrm{co}L_\mathrm{co})
\end{pmatrix}
\end{equation}
where $\overline{L} = (L_\mathrm{A} + L_\mathrm{B})/2$ and $\Delta L = (L_\mathrm{B} - L_\mathrm{A})/2$. Each one of the two eigenvalues defines a transverse mode family of the system. Furthermore, when the accumulated phase $\Theta$ equals an integer multiple of $2\pi$, a longitudinal mode can be found at the corresponding frequency. Diagonalizing the $T$ matrix gives
\begin{equation}
\Theta = n_\mathrm{wg}\omega \overline{L}/c \mp \arccos[\cos(g_\mathrm{co}L_\mathrm{co})\cos(n_\mathrm{wg}\omega \Delta L/c)]
\end{equation}

\begin{figure}
\centering
\includegraphics[width=170mm]{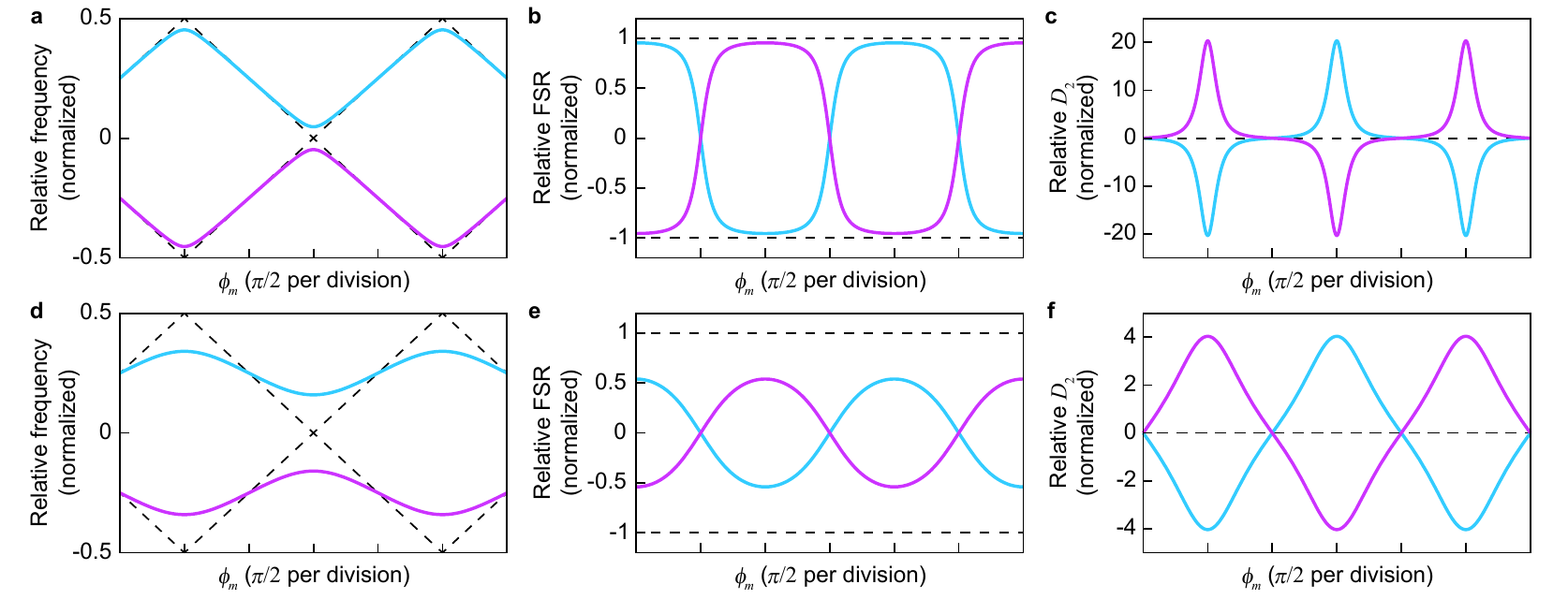}
\captionsetup{singlelinecheck=off, justification = RaggedRight}
\caption{Eigenfrequency plots for the two-ring coupled resonator.
{\bf a}-{\bf c} Relative frequency (normalized to $D_{1,m}$), relative FSR (normalized to $\epsilon D_{1,m}$) and relative $D_2$ (normalized to $\epsilon^2 D_{1,m}$) plots for $g_\mathrm{co}L_\mathrm{co} = 0.3$.
{\bf d}-{\bf f} Similar plots but with $g_\mathrm{co}L_\mathrm{co} = 1.0$. The horizontal axis is defined as $\phi_m = 2\pi \epsilon m$. Relative mode frequency, FSR and $D_2$ for individual rings before coupling have been superimposed (black dashed lines). The relative FSR is found by differentiating the relative frequency, and the relative $D_2$ is found by differentiating the relative FSR.}
\label{FigS2}
\end{figure}

Now we define a mode number associated with the average length of the rings:
\begin{equation}
m \equiv \frac{n_\mathrm{wg}\omega \overline{L}}{2\pi c}
\end{equation}
The relation can be inverted to give a solution of $\omega_m$ dependent on $m$. When $m$ is an integer, $\omega_m$ would be the mode frequencies for a ring resonator with length $\overline{L}$. As $g_\mathrm{co} \ll \omega/c$ and $\Delta L \ll \overline{L}$, the phase contribution related to the coupling varies slowly compared to the $n_\mathrm{wg}\omega \overline{L}/c$ part. This allows us to approximate the coupled mode frequencies using $\omega_m$, and the eigenfrequencies $\omega_{m,\pm}$ can be solved as:
\begin{align}
2\pi m &= n_\mathrm{wg}\omega_{m,\pm} \overline{L}/c \mp \arccos[\cos(g_\mathrm{co}L_\mathrm{co})\cos(n_\mathrm{wg}\omega_{m,\pm} \Delta L/c)] \nonumber\\
&\approx n_\mathrm{wg}\omega_{m,\pm} \overline{L}/c \mp \arccos[\cos(g_\mathrm{co}L_\mathrm{co})\cos(n_\mathrm{wg}\omega_m \Delta L/c)]
\end{align}
\begin{align}
\omega_{m,\pm} &= \omega_m \pm  \left.\left[\frac{\overline{L}}{c}\frac{\partial(n_\mathrm{wg}\omega)}{\partial\omega}\right]^{-1}\right|_{\omega = \omega_m}\times\arccos\left[\cos(g_\mathrm{co}L_\mathrm{co})\cos\left(2\pi \frac{\Delta L}{\overline{L}}m\right)\right] \nonumber\\
&= \omega_m \pm \frac{D_{1,m}}{2\pi}\arccos\left[\cos(g_\mathrm{co}L_\mathrm{co})\cos\left(2\pi \epsilon m\right)\right]
\label{dispersion}
\end{align}
where $D_{1.m}$ is the local FSR that depends on $\omega_m$ and $\epsilon = \Delta L / \overline{L}$ is the length contrast of the rings. The result shows that the mode structure can be seen as splitting off from the length-averaged resonator modes, where the splitting gap is determined by $g_\mathrm{co}$ and modulated with respect to mode number with period $\epsilon^{-1}$.
Note that Eq. (2) in the main text is obtained by replacing the mode number $m$ with the relative mode number $\mu$ in Eq. (\ref{dispersion}).
Such a variable change is valid when $\mu$ is referenced to a frequency degeneracy of the rings.

To gain insight into the model, Fig. \ref{FigS2} plots mode frequency, FSR and the second-order dispersion parameter $D_2$ relative to $\omega_m$ for different values of $g_\mathrm{co}$ as predicted by Eq. (\ref{dispersion}). In these plots $\omega_m$ has been subtracted from the mode frequencies, and only the contributions associated with FSR difference of the two rings and the coupling are considered. The $D_{1,m}$ is also approximated as a constant. The FSRs of the transverse modes show a typical avoided crossing behavior as shown in Fig. \ref{FigS2}b and \ref{FigS2}e. The FSR of one mode continuously transitions to the other mode at the avoided crossing, and similar to the coupling itself, this process is also periodic in the frequency domain. The calculated $D_2$ shows spikes at the avoided crossing center, and the positive spike can be used to counter the normal dispersion present in the averaged resonator dispersion. Smaller $g_\mathrm{co}$ leads to higher peak $D_2$ with smaller crossing bandwidth. To get a larger crossing bandwidth, $g_\mathrm{co}$ could be increased at the expense of lower $D_2$, but the maximum bandwidth is half the modulation period (i.e. the vernier FSR) as the effect of the neighboring crossings set in and shifts the $D_2$ in the opposite direction.

\begin{figure}
\centering
\includegraphics[width=170mm]{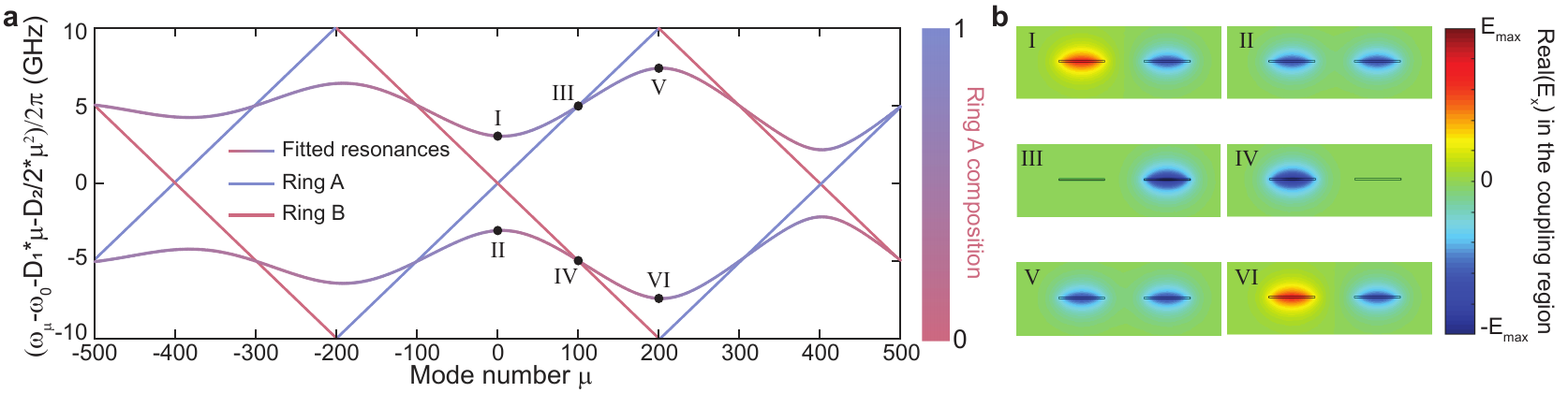}
\captionsetup{singlelinecheck=off, justification = RaggedRight}
\caption{Illustration of mode hybridization in the coupling region. This figure reproduces Extended Data Fig. 2 from the main text.
\textbf{a} Fitted optical resonance frequency dispersion of the coupled resonator (solid curves) and fitted mode frequency dispersion of the single rings (red and blue lines) plotted versus relative mode number $\mu$.
These plots are the same as Fig. 1e in the main text.
\textbf{b} Cross-sectional view of simulated electric field amplitudes in the coupled region at mode numbers indicated in panel \textbf{a} by the black points. The right (left) waveguide belongs to ring A (B). At the crossing center (I, II, V and VI), two waveguides have the same field intensity and the opposite (same) phase for the anti-symmetric (symmetric) mode. When hybrid mode frequencies meet the single-ring resonances (III and IV), the electrical field at the coupled region is contributed by a single ring.
}
\label{FigS3}
\end{figure}

In addition to the mode frequency, the mode compositions can also be derived from the transfer matrix $T$. As the change of mode profile is large enough across the measured optical bandwidth, the mode compositions has an impact on soliton power distribution in the rings (as in Fig. 2a in the main text), and complements FSRs and dispersions when describing the dispersion characteristics. The eigenvectors of $T$ read,
\begin{equation}
\psi_\mathrm{I} \propto \left(\sqrt{\frac{\sin(\alpha + \phi_m)}{2\sin\alpha\cos\phi_m}},\ \  \mp\sqrt{\frac{\sin(\alpha - \phi_m)}{2\sin\alpha\cos\phi_m}}\right)^\mathbf{T},\ \ 
\phi_m = 2\pi \epsilon m,\ \ 
\alpha = \mp\arccos\left[\cos(g_\mathrm{co}L_\mathrm{co})\cos\phi_m\right]
\end{equation}
This gives the relative field intensities in the non-coupled regions of the rings for a particular mode, and is used to plot composition ratios in Fig. 1d, Fig. 1e, and Fig. 2c in the main text. Another point of interest is the center of the coupled region. Here the field can be found as
\begin{equation}
\psi_\mathrm{co} \propto
\begin{pmatrix}
\cos(g_\mathrm{co}L_\mathrm{co}/2) & i\sin(g_\mathrm{co}L_\mathrm{co}/2) \\
i\sin(g_\mathrm{co}L_\mathrm{co}/2) & \cos(g_\mathrm{co}L_\mathrm{co}/2)
\end{pmatrix}
\begin{pmatrix}
e^{i\phi_m/2} & 0 \\
0 & e^{-i\phi_m/2}
\end{pmatrix}
\psi_\mathrm{I} \propto
\left(\sqrt{\frac{\sin\alpha + \sin\phi_m}{2\sin\alpha}},\ \  \mp\sqrt{\frac{\sin\alpha - \sin\phi_m}{2\sin\alpha}}\right)^\mathbf{T}
\end{equation}
There are some special cases of $\phi_m$ that lead to simplified field distributions and are demonstrated in Fig. \ref{FigS3}. For example, if $\phi_m/\pi$ is an integer (crossing centers), the modes become purely symmetrical and anti-symmetrical:
\begin{equation}
\psi_\mathrm{I}\propto (\sqrt{1/2}, \mp\sqrt{1/2})^\mathbf{T}, \ \ 
\psi_\mathrm{co}\propto (\sqrt{1/2}, \mp\sqrt{1/2})^\mathbf{T}, \ \ 
\end{equation}
Points I, II, V, and VI in Fig. \ref{FigS3}b belong to these cases. Points II and V are symmetric modes formed by the two rings, with equal mode intensities and the same phase. On the other hand, points I and VI are anti-symmetric modes, with equal mode intensities but opposite phase. These results happen to agree with coupled-mode calculations when only the pair of degenerate longitudinal modes from each ring are considered. However, while the energy is equally distributed in the two rings in the same way as the reduced coupled-mode theory predicted, other longitudinal modes still participate in the coupling because the wavevector in the coupled region differs from that in the uncoupled region. On the other hand, if $\phi_m/\pi$ is a half-integer (halfway between crossing centers), then at the center of coupling position the field is entirely within a single ring:
\begin{equation}
\psi_\mathrm{I}\propto (\cos(g_\mathrm{co}L_\mathrm{co}/2), \sin(g_\mathrm{co}L_\mathrm{co}/2))^\mathbf{T}, \ \ 
\psi_\mathrm{co}\propto (1, 0)^\mathbf{T}, \ \ 
\text{or} \ \ 
\psi_\mathrm{I}\propto (\sin(g_\mathrm{co}L_\mathrm{co}/2), -\cos(g_\mathrm{co}L_\mathrm{co}/2))^\mathbf{T}, \ \ 
\psi_\mathrm{co}\propto (0, -1)^\mathbf{T}, \ \ 
\end{equation}
Points III and IV in Fig. \ref{FigS3}b belong to these cases. 

An interesting feature of the field distribution is that, for a single continuous branch, the field compositions exchange parity at the next degeneracy point, and the antisymmetric mode now becomes the symmetric mode (from point I to V) and vice versa (from point II to VI). The change of parity shows that the modes repeat themselves every two vernier periods (every two degeneracy points) instead of one, in agreement with Eq. (\ref{dispersion}). While the parity exchange is obvious after plotting the dispersion (Fig. \ref{FigS3}a), it can also be understood from a mode number argument. We consider the total phase accumulated in ring A for a specific mode divided by $2\pi$, which should be an integer and denoted as $m_\mathrm{A}$. This is the ``mode number'' for ring A for the specific mode. Similarly $m_\mathrm{B}$ could be defined. These two numbers equal to the respective mode numbers of the closest uncoupled modes, which can be seen by adiabatically turning of the coupling. For a single vernier period, the total mode number changes by an odd number. However, going to the next longitudinal mode by changing the frequency alone changes both $m_\mathrm{A}$ and $m_\mathrm{B}$ by one. The only way to induce a separate mode number change is to create a zero in the field amplitude somewhere in the respective ring, which is indeed the case for points III and IV shown in Fig. \ref{FigS3}b. Considering that the individual mode numbers are about equally distributed around the averaged-length mode number $m$ (e.g. $|(m_\mathrm{B}-m)-(m-m_\mathrm{A})|\leq 1$), the extra increment of $m_\mathrm{B}$ and decrement of $m_\mathrm{A}$ should have taken place alternatively between the vernier periods, indicating the mode branch switches mode compositions for each vernier period.

\medskip

\section{Eigenmodes of a three-ring coupled resonator}

In this section we study the mode frequencies of the three-ring coupled resonator. Although the derivation is similar to that of the two-ring resonator, we will highlight some features of the coupled system that are not obvious in the two-ring case. The result can also be readily generalized to multi-ring arrangements.

\begin{figure}[h!]
\centering
\includegraphics[width=170mm]{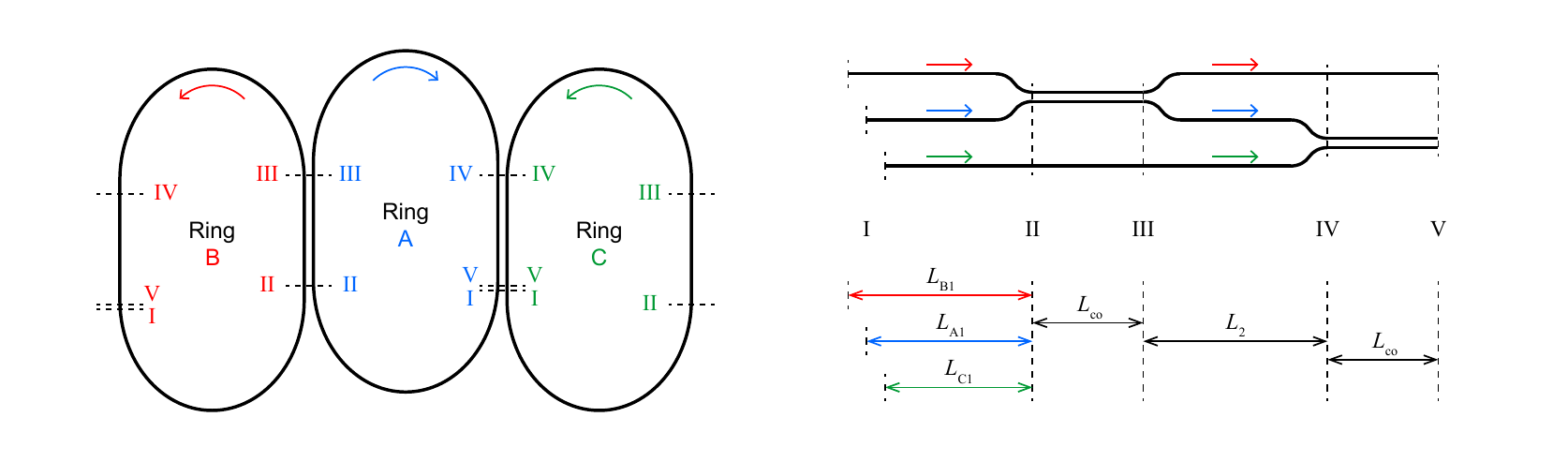}
\captionsetup{singlelinecheck=off, justification = RaggedRight}
\caption{Schematic of the three-ring coupled resonator.
Left panel: Top view of the coupled resonator with key points marked.
Right panel: Schematic of the resonator with straightened waveguides (not to scale). Segment lengths have been marked.}
\label{FigS6}
\end{figure}

The schematic for the three-ring coupled resonator is shown in Fig. \ref{FigS6} along with definitions of segment lengths. Unlike the two-ring case, there is no explicit symmetry to take advantage of, and the segmentation method is chosen to reduce calculation complexity. We can assemble the field amplitudes from ring C, A and B, in that order, into a vector and find the transfer matrix for each section:
\begin{equation}
\psi_\mathrm{II}
=
\exp{\left[i n_\mathrm{wg}\omega/c
\begin{pmatrix}
L_\mathrm{C1} & 0 & 0 \\
0 & L_\mathrm{A1} & 0 \\
0 & 0 & L_\mathrm{B1}
\end{pmatrix}
\right]}
\psi_\mathrm{I}
\end{equation}

\begin{equation}
\psi_\mathrm{III}
=
\exp{\left[i L_\mathrm{co}
\begin{pmatrix}
n_\mathrm{wg}\omega/c & 0 & 0 \\
0 & n_\mathrm{wg}\omega/c & g_\mathrm{co} \\
0 & g_\mathrm{co} & n_\mathrm{wg}\omega/c
\end{pmatrix}
\right]}
\psi_\mathrm{II}
\end{equation}

\begin{equation}
\psi_\mathrm{IV}
=
\exp{\left[i n_\mathrm{wg}\omega/c
\begin{pmatrix}
L_\mathrm{2} & 0 & 0 \\
0 & L_\mathrm{2} & 0 \\
0 & 0 & L_\mathrm{2}
\end{pmatrix}
\right]}
\psi_\mathrm{III}
\end{equation}

\begin{equation}
\psi_\mathrm{V}
=
\exp{\left[i L_\mathrm{co}
\begin{pmatrix}
n_\mathrm{wg}\omega/c & g_\mathrm{co} & 0 \\
g_\mathrm{co} & n_\mathrm{wg}\omega/c & 0 \\
0 & 0 & n_\mathrm{wg}\omega/c
\end{pmatrix}
\right]}
\psi_\mathrm{IV}
\end{equation}

The overall round-trip transfer matrix is the product of the previous four matrices and reads
\begin{equation}
T=
\begin{pmatrix}
  e^{in_\mathrm{wg}\omega L_\mathrm{C}/c}\cos(g_\mathrm{co}L_\mathrm{co}) &
i e^{in_\mathrm{wg}\omega L_\mathrm{A}/c}\cos(g_\mathrm{co}L_\mathrm{co})\sin(g_\mathrm{co}L_\mathrm{co}) &
- e^{in_\mathrm{wg}\omega L_\mathrm{B}/c}\sin^2(g_\mathrm{co}L_\mathrm{co}) \\
i e^{in_\mathrm{wg}\omega L_\mathrm{C}/c}\sin(g_\mathrm{co}L_\mathrm{co}) &
  e^{in_\mathrm{wg}\omega L_\mathrm{A}/c}\cos^2(g_\mathrm{co}L_\mathrm{co}) & 
i e^{in_\mathrm{wg}\omega L_\mathrm{B}/c}\cos(g_\mathrm{co}L_\mathrm{co})\sin(g_\mathrm{co}L_\mathrm{co}) \\
0 &
i e^{in_\mathrm{wg}\omega L_\mathrm{A}/c}\sin(g_\mathrm{co}L_\mathrm{co}) &
  e^{in_\mathrm{wg}\omega L_\mathrm{B}/c}\cos(g_\mathrm{co}L_\mathrm{co})
\end{pmatrix}
\end{equation}
where we defined the total length of ring C, $L_\mathrm{C}=L_\mathrm{C1}+L_2+2L_\mathrm{co}$, and $L_\mathrm{A}$ and $L_\mathrm{B}$ are defined similarly. Note that the dependence on individual segment lengths $L_\mathrm{C1}$, $L_\mathrm{A1}$, $L_\mathrm{B1}$ and $L_2$ have disappeared from $T$, indicating that the relative position of the couplers on ring A does not matter for eigenfrequency calculations. This is because propagating the same distance for all three components provides only a global phase for the state, which can be moved past the coupler. Mathematically, the coupling matrix commutes with the propagation matrix, which is proportional to the identity matrix for identical ring cross sections:
\begin{equation}
\left[
L_\mathrm{co}
\begin{pmatrix}
n_\mathrm{wg}\omega/c & 0 & 0 \\
0 & n_\mathrm{wg}\omega/c & g_\mathrm{co} \\
0 & g_\mathrm{co} & n_\mathrm{wg}\omega/c
\end{pmatrix},\ \ 
n_\mathrm{wg}\omega/c
\begin{pmatrix}
L_\mathrm{2} & 0 & 0 \\
0 & L_\mathrm{2} & 0 \\
0 & 0 & L_\mathrm{2}
\end{pmatrix}
\right]
=0
\end{equation}
For the coupler itself, the propagating part (diagonal elements) also commute with the pure coupling part (off-diagonal elements), although different couplers do not commute. Therefore, the system is equivalent to propagating along the entire length of individual rings, followed by two point couplers with the same coupling ratios as the original couplers. This argument works for all coupled resonators with identical ring cross-sections coupled in a chain or tree topology, and provides a degree of freedom for placing the rings in the design phase.

\begin{figure}
\centering
\includegraphics[width=170mm]{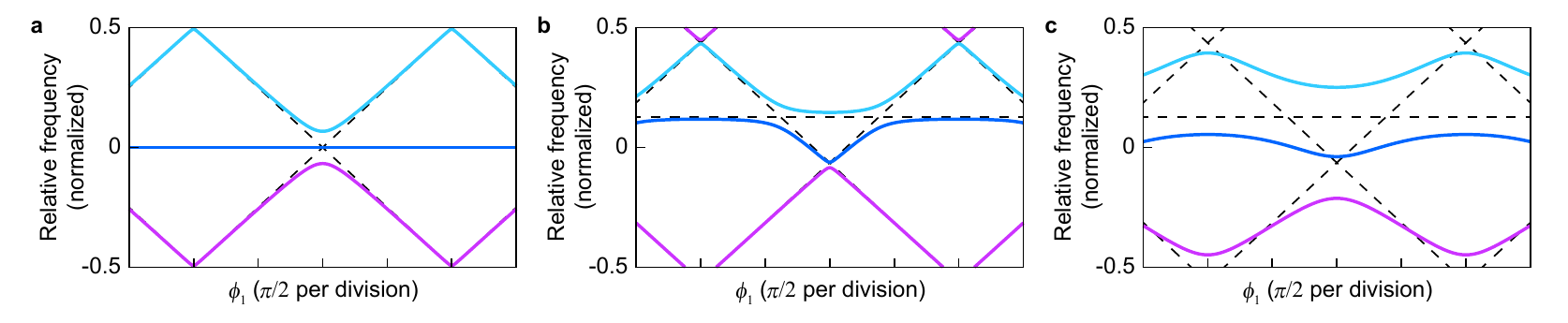}
\captionsetup{singlelinecheck=off, justification = RaggedRight}
\caption{Eigenfrequency plots for the three-ring coupled resonator, showing relative frequency (normalized to $D_{1,m}$) versus $\phi_1$. Parameters are
{\bf a} $g_\mathrm{co}L_\mathrm{co} = 0.3$ and $\phi_2 = 0$;
{\bf b} $g_\mathrm{co}L_\mathrm{co} = 0.3$ and $\phi_2 = 0.4$;
{\bf c} $g_\mathrm{co}L_\mathrm{co} = 1.0$ and $\phi_2 = 0.4$.}
\label{FigS7}
\end{figure}

Following the two-ring analysis, we define an averaged length for the resonators and its associated mode number:
\begin{equation}
\overline{L} \equiv \frac{L_\mathrm{C}+L_\mathrm{A}+L_\mathrm{B}}{3},\ \ 
m \equiv \frac{n_\mathrm{wg}\omega \overline{L}}{2\pi c}
\end{equation}
We will also need to define two length differences. For the current design, we have $L_\mathrm{C} + L_\mathrm{B} \approx 2L_\mathrm{A}$, and the following contrast definitions become convenient:
\begin{equation}
\epsilon_1 = \frac{L_\mathrm{B}-L_\mathrm{C}}{2\overline{L}},\ \  \epsilon_2 = \frac{L_\mathrm{C}+L_\mathrm{B}-2L_\mathrm{A}}{6\overline{L}}
\end{equation}
With these notations, $T$ can be written as
\begin{equation}
T=e^{2\pi m i}
\begin{pmatrix}
  e^{i(-\phi_1+\phi_2)}\cos(g_\mathrm{co}L_\mathrm{co}) &
i e^{-2i\phi_2}\cos(g_\mathrm{co}L_\mathrm{co})\sin(g_\mathrm{co}L_\mathrm{co}) &
- e^{i(\phi_1+\phi_2)}\sin^2(g_\mathrm{co}L_\mathrm{co}) \\
i e^{i(-\phi_1+\phi_2)}\sin(g_\mathrm{co}L_\mathrm{co}) &
  e^{-2i\phi_2}\cos^2(g_\mathrm{co}L_\mathrm{co}) & 
i e^{i(\phi_1+\phi_2)}\cos(g_\mathrm{co}L_\mathrm{co})\sin(g_\mathrm{co}L_\mathrm{co}) \\
0 &
i e^{-2i\phi_2}\sin(g_\mathrm{co}L_\mathrm{co}) &
  e^{i(\phi_1+\phi_2)}\cos(g_\mathrm{co}L_\mathrm{co})
\end{pmatrix}
\end{equation}
with $\phi_1 = 2\pi\epsilon_1 m$ and $\phi_2 = 2\pi\epsilon_2 m$. For the current design, $\epsilon_1\approx 3\times 10^{-3}$ and $\epsilon_2\approx 1.5\times 10^{-6}$, which ensures a slowly-varying phase contributed by the coupling. The eigenfrequencies are given by
\begin{equation}
\omega = \omega_m - \frac{D_{1,m}}{2\pi}\theta
\end{equation}
where $e^{i\theta}$ is given by the roots to the cubic characteristic equation:
\begin{equation}
x^3
-(e^{-2i\phi_2}\cos(g_\mathrm{co}L_\mathrm{co})+2e^{i\phi_2}\cos(\phi_1))\cos(g_\mathrm{co}L_\mathrm{co})x^2
+(e^{2i\phi_2}\cos(g_\mathrm{co}L_\mathrm{co})+2e^{-i\phi_2}\cos(\phi_1))\cos(g_\mathrm{co}L_\mathrm{co})x
-1=0,\ \ x \equiv e^{i\theta}
\end{equation}
The unitary nature of $T$ ensures that all three roots for $x$ lie on the complex unit circle.

As $\epsilon_2 \ll \epsilon_1$ for the current design, $\phi_2$ varies much more slowly compared to $\phi_1$ and we will take $\phi_2$ to be a constant to simplify the discussions below. Fig. \ref{FigS7} plots the relative frequencies for some parameter combinations. In the case of $\phi_2 = 0$ (Fig. \ref{FigS7}a), the mode frequencies of ring A coincides with the averaged frequency, and the mode frequencies of ring A and B are symmetrically distributed around the averaged frequency. As a result, the coupled frequency spectrum resembles that of the two-ring resonator. A key difference here is that the two gaps opened have different widths. For crossings at integer $\phi_1/(2\pi)$ locations, ring A participates in the coupling, and the total gap is approximately $2\sqrt{2}g_\mathrm{co}L_\mathrm{co}\times D_{1,m}/(2\pi)$ for small $g_\mathrm{co}L_\mathrm{co}$. For the other crossings at half-integer $\phi_1/(2\pi)$ locations, the mode from ring A is half an FSR away from ring C and B, and the coupling becomes indirect. Here the gap width is approximately $(g_\mathrm{co}L_\mathrm{co})^2\times D_{1,m}/(2\pi)$ for small $g_\mathrm{co}L_\mathrm{co}$, which is second order in the coupling strength. A nonzero $\phi_2 = 0$ breaks the frequency-domain symmetry and leads to additional avoided crossings (Fig. \ref{FigS7}b). For stronger coupling strengths, the bandwidths of the crossings expand and merge with the other crossings (Fig. \ref{FigS7}c) similar to the two-ring case. Here the frequency dispersion become smoother and have less overall coupling-contributed dispersion.

\begin{figure}
\centering
\includegraphics[width=170mm]{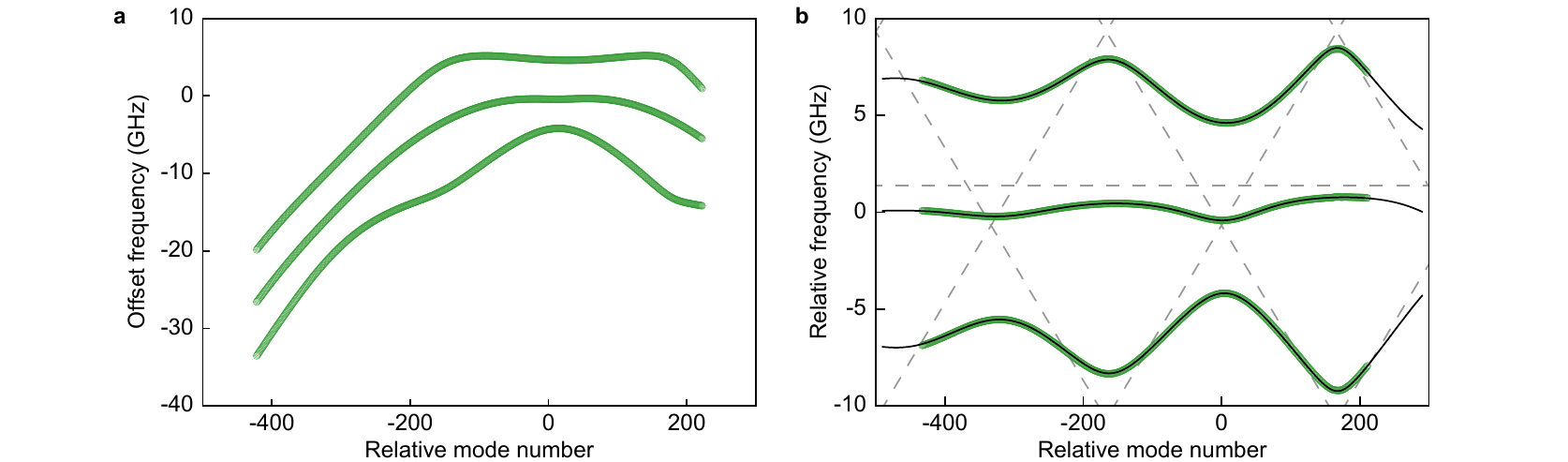}
\captionsetup{singlelinecheck=off, justification = RaggedRight}
\caption{Dispersion of the three-ring resonator.
{\bf a} Measured dispersion for the three-ring resonator. The topmost branch is also shown in Fig. 4c in the main text.
{\bf b} Mode frequencies relative to the averaged frequencies (circles). Solid curves show the fitted result using the three-ring model, and are in excellent agreement with the data. Dashed lines indicate frequencies of the individual rings before coupling.}
\label{FigS8}
\end{figure}

Figure \ref{FigS8}a shows the measured dispersion data for the three-ring resonator. The topmost branch is used for soliton generation and is also shown in Fig. 4c in the main text. Fitting the averaged frequency (not plotted) gives an averaged FSR of 19.9711 GHz and a second-order dispersion parameter of 282.7 kHz, consistent with the two-ring results. After subtracting the averaged frequency, Fig. \ref{FigS8}b shows the relative frequency with a similar structure of Fig. \ref{FigS7}c. Using the exponential decaying coupling model as described in the methods ($g_\mathrm{co} = g_\mathrm{co,0}\exp(-\mu/\mu_g)$), the fitted result for the relative frequencies also shows good agreement with the measured data. The fitted parameters are $g_\mathrm{co,0}L_\mathrm{co} = 0.985$, $\mu_g = 1175$ and $\phi_2 = 0.216$.

\end{document}